\documentclass[10pt,journal,final]{IEEEtran}

\makeatletter

\newcommand{\Rmnum}[1]{\expandafter\@slowromancap\romannumeral #1@}
\makeatother
\usepackage{bm}
\usepackage{multirow} 
\usepackage{booktabs}
\usepackage{makecell}
\usepackage{algorithm}   
\usepackage{rotating}
\usepackage{algpseudocode}
\usepackage{mathrsfs}

\usepackage{subfigure}
\usepackage{setspace}

\newcommand{\RNum}[1]{\uppercase\expandafter{\romannumeral #1\relax}}
\usepackage{cite}
\usepackage{graphicx}
\graphicspath{{./figures/}}
\usepackage{balance}
\usepackage{float}
\usepackage{amssymb}
\usepackage{amsthm}
\usepackage{amsmath}
\usepackage{color}

\ifCLASSOPTIONcompsoc
\usepackage[caption=false,font=normalsize,labelfont=sf,textfont=sf]{subfig}
\else
\usepackage[caption=false,font=footnotesize]{subfig}
\fi
\begin{document}	
\title{Learning an Interpretable End-to-End Network for Real-Time Acoustic Beamforming
	{\footnotesize}
	\thanks{The work was supported in part by National Natural Science Foundation of China under Grants 52105126, 82172033, 82272071, and the China Fundamental Research Funds for the Central Universities (Grant No. 20720220082). (*Corresponding author: Xiaotong Tu; Electronic mail: xttu@xmu.edu.cn)
	
	Hao Liang, Guanxing Zhou, and Xiaotong Tu are with the School of Informatics, Xiamen University, Xiamen 361005, China (E-mail: haoliang@stu.xmu.edu.cn; joaquinchou@stu.xmu.edu.cn; xttu@xmu.edu.cn).
	
	Andreas Jakobsson is with the Centre for Mathematical Sciences, Lund University, Sweden (E-mail: aj@maths.lth.se).
	
	Xinghao Ding and Yue Huang are with the Institue of Artificial Intelligent, Xiamen University, Xiamen 361005, China (E-mail: dxh@xmu.edu.cn; yhuang2010@xmu.edu.cn).
} 
} 
\author{Hao Liang, Guanxing Zhou, Xiaotong Tu*, Andreas Jakobsson, Xinghao Ding, Yue Huang}

\maketitle
\begin{abstract}
Recently, many forms of audio industrial applications, such as sound monitoring and source localization, have begun exploiting smart multi-modal devices equipped with a microphone array. Regrettably, model-based methods are often difficult to employ for such devices due to their high computational complexity, as well as the difficulty of appropriately selecting the user-determined parameters. As an alternative, one may use deep network-based methods, but these are often difficult to generalize, nor can they generate the desired beamforming map directly. In this paper, a computationally efficient acoustic beamforming algorithm is proposed, which may be unrolled to form a model-based deep learning network for real-time imaging, here termed the DAMAS-FISTA-Net. By exploiting the natural structure of an acoustic beamformer, the proposed network inherits the physical knowledge of the acoustic system, and thus learns the underlying physical properties of the propagation. As a result, all the network parameters may be learned end-to-end, guided by a model-based prior using back-propagation. Notably, the proposed network enables an excellent interpretability and the ability of being able to process the raw data directly. Extensive numerical experiments using both simulated and real-world data illustrate the preferable performance of the DAMAS-FISTA-Net as compared to alternative approaches.
\end{abstract}


\IEEEpeerreviewmaketitle

\section{Introduction}
\begin{figure}[!t]
	\centering
	\includegraphics[width=\linewidth]{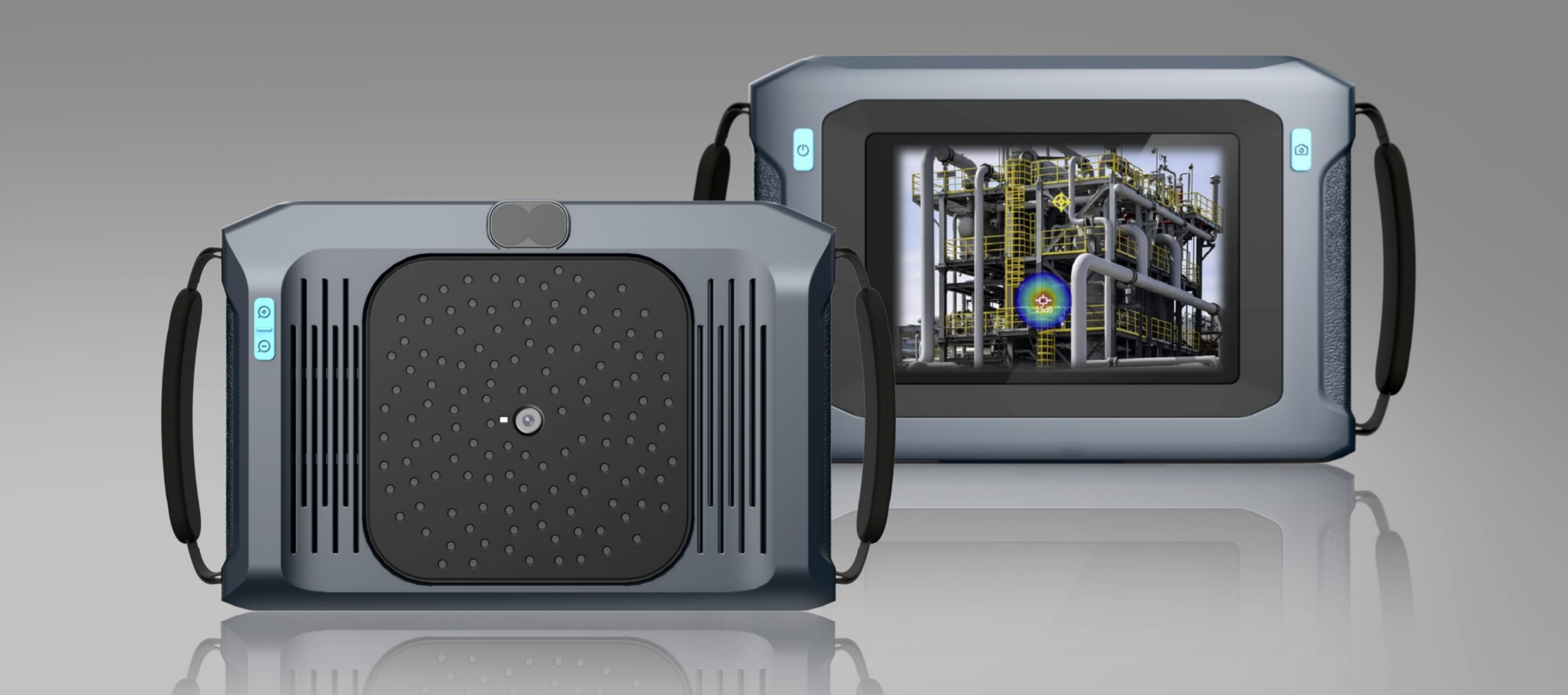}
	\caption{Schematic of an acoustic camera. The light spot indicates the location and power of an industrial anomalous sound source.}
	\label{fig1}
\end{figure}

\IEEEPARstart{T}he acoustic camera, a multi-modal imaging device, which can visualize sound as a heatmap, has been widely applied in various fields, such as transportation, noise monitoring, and industrial anomaly detection \cite{wall2012spiking,savitha2013projection,davila2018enhanced,salvati2020diagonal,songgong2022acoustic,sun2022beyond}. As illustrated of such a device in Fig.\ref{fig1}, where the acoustic camera is able to make sound ``visible". It processes the multi-dimension signals acquired by the microphone array, employing acoustic beamforming algorithms to generate the sound pressure level distribution on the scanning plane, and displays the results as images or videos. Utilizing the phase-shifting properties of the signal propagation, the delay-and-sum (DAS) algorithm provides a robust, fast, and intuitive imaging result via delay compensation and weighted summation \cite{johnson1992array,van1988beamforming}. However, it is subject to the Sparrow resolution limit \cite{sparrow1916spectroscopic}, and thus yields low-resolution results with high sidelobes \cite{merino2019review}.

In order to eliminate these side-lobe interferences, various deconvolution approaches have been proposed. In \cite{hogbom1974aperture}, the CLEAN-PSF method was introduced to iteratively remove the influence of the point spread function (PSF). Then, based on the spatial source coherence, Sijtsma further extended this method to allow for coherent sources, resulting in the so-called CLEAN-SC algorithm \cite{sijtsma2007clean}. Later, a high-resolution extension of CLEAN-SC was developed in \cite{sijtsma2017high}. However, the CLEAN-based methods may erroneously "clean" the true source point, and the deconvolution may as a result become unfocused in the acoustic field. Assuming that the beamforming map is a linear combination of the sources and the PSF, the deconvolution approach for the mapping of incoherent acoustic sources (DAMAS), which removes the PSF's influence by solving an inverse problem, was introduced in \cite{brooks2006deconvolution}. Building on this, a series of DAMAS-based methods exploiting some structured sparsity and intrinsic acoustic properties have been derived \cite{dougherty2005extensions,brooks2006extension,ehrenfried2007comparison,lylloff2015improving,herold2017comparison,ding2022high}, all aiming to form higher quality and more robust estimates. Although these methods enable a significant improvement of the results, they usually require a high computational cost, preventing their usage in real-time imaging, as well as necessitating an often difficult selection of various user-defined parameters.

Recently, several deep network-based acoustic beamforming approaches have also been proposed to reconstruct acoustic sources \cite{reiter2017machine,allman2018photoacoustic,kujawski2019deep,zhou2022acoustic}. Compared with the model-based procedures, these methods can dramatically reduce the necessary time complexity, while providing reliable estimates as a result of the strong feature extraction capability of the neural network. However, most existing deep network-based algorithms tend to train the neural network as a black box, inevitably leading to the absence of domain knowledge guidance, which often causes the resulting network to be difficult to generalize in a robust manner, limiting the practical application of these methods.

In this work, we propose a computationally efficient implementation of the DAMAS, termed the DAMAS-FISTA, which employs the Fast Iterative Shrinkage Thresholding Algorithm (FISTA) \cite{beck2009fast}. Using this, we then design an interpretable end-to-end network, dubbed the DAMAS-FISTA-Net, by mapping the DAMAS-FISTA to render a high-quality beamforming map for real-time imaging. In contrast to most deep network architectures, the proposed network can directly handle the raw data without pre-beamforming. Notably, the proposed DAMAS-FISTA-Net can also learn the underlying physical properties of the acoustic environment by exploiting the model-based method, allowing the resulting network an excellent ability to be generalized. All the required network parameters may be learned end-to-end, guided by the model-based prior, by using back-propagation. Numerical simulations and experimental data analysis show the effectiveness and advantages of the proposed method.

In summary, the main contributions of this paper are:
\begin{itemize}
	\item We introduce a computationally efficient implementation of the DAMAS estimator by employing a FISTA framework, forming the so-called DAMAS-FISTA method, which may then be used to guide the design of the proposed subsequent network.
	\item We propose the novel and computationally efficient DAMAS-FISTA-Net, which not only combines the benefit of the model-based method with those of deep learning, but which inherits related domain knowledge.
	\item The proposed network exhibits an excellent interpretability and ability to be generalized, while still being capable of processing the raw data directly.
	\item Extensive experiments illustrate that the DAMAS-FISTA-Net achieves promising performance for both simulated and real-world data, illustrating the practical potential of the method.
\end{itemize}

The remainder of this paper is organized as follows: In the next section, related work, including several acoustic beamforming methods and model-based deep learning methods, is summarized. Then, Section \ref{s3} presents the acoustic signal model and some related notations, and the DAMAS beamformer is briefly reviewed. In Section \ref{s4}, the DAMAS-FISTA algorithm is proposed, followed by the proposed DAMAS-FISTA-Net method. Section \ref{s5} provides a comparison and validation of the proposed methods, formed by analyzing both simulated and real-world signals. Finally, our conclusions are summarized in Section \ref{s6}.

\section{Related Work}

There are two main categories of related acoustic beamforming methods, including model-based methods, such as the deconvolution-based algorithms, and deep network-based methods. A brief review of these two approaches is found below. Additionally, we provide a concise review of the recent model-based deep learning approaches which are relevant to our work.

\textbf{Model-based acoustic beamforming.}
In the field of acoustic beamforming, the DAMAS algorithm introduced the use of deconvolution to increase the resolution and remove disturbing side-lobe interference \cite{brooks2006deconvolution}. However, this algorithm is unsuitable for coherence sources and suffers from being computationally cumbersome limitation. Notable efforts have been made to address these issues. In \cite{brooks2006extension}, the cross-power between coherent sources have been considered, while the conjugate relationship is used to reduce the computational complexity. However, computationally intensive operations are still unavoidable. Assuming that the PSF is shift-invariant, Dougherty proposed the DAMAS2, significantly improving the computational efficiency \cite{dougherty2005extensions}. Later, the computational overhead was further reduced by using efficient solvers \cite{ehrenfried2007comparison,lylloff2015improving}. More recently, Ding \emph{et al.} proposed two high-resolution source localization methods, which exploit the sparsity of the beamforming map \cite{ding2022high}. Although these methods perform well, a degree of shift-variance is introduced due to the shift-invariant assumption, casuing the source power to be underestimated. Generally, all these model-based methods require hundreds or thousands of iterations even when using efficient solvers, which will inevitably increase the computational burden and restrict the application for real-time imaging. Additionally, many of these algorithms depend on a series of user-defined parameters, such as sparity or penalty parameters, which typically are non-trivial to select in an optimal manner, but which will greatly influence the accuracy of the imaging results.

\textbf{Deep network-based acoustic beamforming.}
Recently, many studies have shown that a deep network may be a useful tool for acoustic beamforming by taking advantage of its powerful learning capability. For instance, Reiter and Bell employed the widely used AlexNet \cite{NIPS2012_c399862d} to identify point source locations with its input being the pre-beamformed result \cite{reiter2017machine}. Later, Allman \emph{et al.} treated the source localization problem as a classification task \cite{allman2018photoacoustic}, designing a two-module deep network based on the VGGNet \cite{simonyan2014very} and Fast R-CNN \cite{girshick2015fast}. However, these two methods only estimate the localization information, without providing any intensity prediction. To handle this issue, Kujawski \emph{et al.} proposed to use the ResNet architecture \cite{he2016deep} to process the pre-beamformed result to estimate the position as well as the intensity of a single source \cite{kujawski2019deep}. To avoid information loss during pre-beamforming, Zhou \emph{et al.} further proposed the Acoustic-Net \cite{zhou2022acoustic}, which employs the time-frequency distribution of multi-dimension signals as input. Combined with the RepVGG architecture \cite{ding2019acnet}, the inference time can then be significantly reduced. However, the above networks are only suitable for single-source situations. Indeed, some recent works (see, e.g., \cite{lee2021deep,feng2022double}) may be suitable for multi-source situations. Unfortunately, to be efficient, these methods generally need notable prior knowledge, such as the number of sources. The main feature of deep network-based methods is that they are completely data-driven, without employing any physical mechanism, resulting in a network that is data-sensitive and lack structural diversity (repetitively employing fully-connected or convolutional layers). Additionally, neither of these methods can directly generate beamforming maps.

\textbf{Model-based deep learning.}
In order to build a bridge between model-based methods and deep network-based methods, the model-based deep learning framework has also been investigated and has attracted notable attention. A representative method is the learned iterative shrinkage and thresholding algorithm (LISTA) \cite{gregor2010learning}, which unrolls the iterative shrinkage and thresholding (ISTA) algorithm \cite{daubechies2004iterative} as a special network to learn fast approximations of sparse coding. Later, Wang \emph{et al.} developed a deep $l_0$ encoder to solve $l_0$ sparse regularization problems \cite{wang2016learning}. Furthermore, Yang \emph{et al.} considered a more generalized compressive sensing problem, resulting in the ADMM-Net \cite{DBLP:conf/nips/Yang0LX16,yang2018admm}. Also, a recurrent neural network, dubbed DeepWave, has been designed for three-dimension imaging \cite{NEURIPS2019_e9bf14a4}. In \cite{zhang2018ista}, an optimization-inspired ISTA-Net was proposed, which may be seen as an extension of LISTA. More recently, Xiang \emph{et al.} further proposed the FISTA-Net by introducing a momentum module and a parameter tweaking strategy \cite{xiang2021fista}. Essentially, the proposed DAMAS-FISTA may be viewed as a special extension of the FISTA-Net, which also provides insights into the acoustic field. Notably, in comparison with the utilization of deep convolutional networks in the FISTA-Net, the proposed network utilizes fewer although significant parameters detailing the natural structure, which thereby effectively avoid the overfitting problem.

\begin{figure}[!t]
	\centering
	\includegraphics[width=\linewidth]{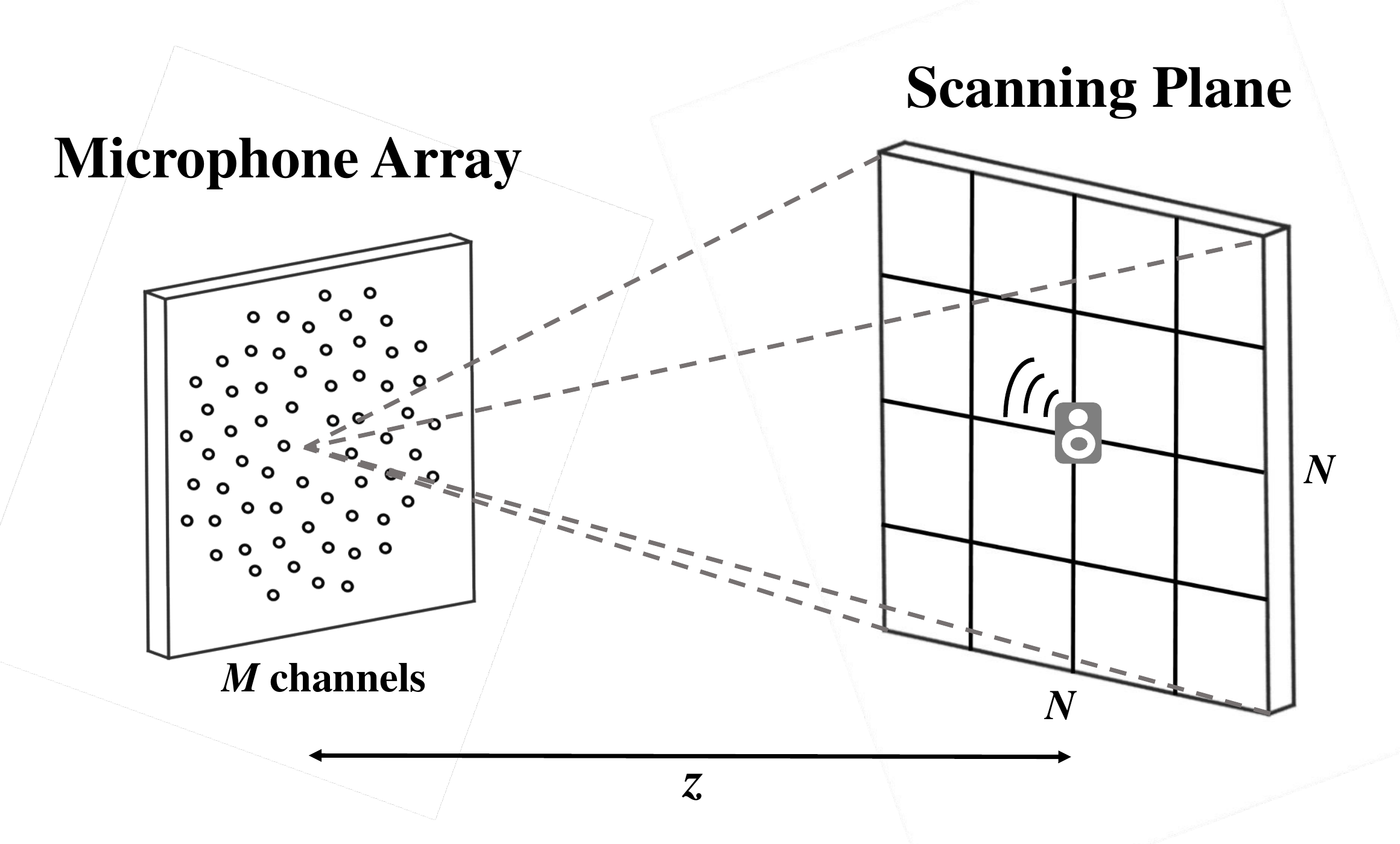}
	\caption{The assumed model of the acoustic signal propagation.}
	\label{fig2}
\end{figure}

\section{Preliminaries} \label{s3}

\subsection{Traditional Beamforming}

The traditional DAS beamforming algorithm exploits the phase difference generated by the relative time-delay on each microphone channel to synchronize the multi-dimensional acoustic signals \cite{johnson1992array,van1988beamforming}. Specifically, as shown in Fig.\ref{fig2}, the signals resulting from unknown sources are measured by a microphone array with $M$ channels, with its scanning plane at a distance of $z$ m from the scanning plane being divided into $N\times N$ grid points. As a delay in the time domain equals a phase shift in the frequency domain, one may organize these delays into a vector, termed the steering vector, which, for the $n$-th grid point, may be expressed as \cite{merino2019review}
\begin{equation} \label{eq1}
	\small
	\bm{g}_n=\left[\begin{array}{ccc}
		\frac{r_0}{r_{1, n}} \mathrm{e}^{-j k\left(r_{1, n}-r_{0}\right)}  & \ldots & \frac{r_0}{r_{M, n}} \mathrm{e}^{-j k\left(r_{M, n}-r_{0}\right)}
	\end{array}\right]^{T},
\end{equation}
where $r_0$ denotes the distance from the array center to the $n$-th grid point, where $r_{m, n}$ is the distance from the $m$-th microphone to the $n$-th grid point, with $k$ representing the wave number, i.e., 
\begin{equation}
	k = \frac{2\pi f}{c},
\end{equation}
where $f$ denotes the scanning frequency and $c$ the speed of sound. A further discussion about the formulations of the steering vector is available in \cite{sarradj2012three}. According to the definition of the steering vector in (\ref{eq1}), the weighted steering vector, for the $n$-th grid point, may be expressed as
\begin{equation} \label{eq3}
	\small
	\begin{aligned}
		\bm{w}_n &= \frac{\bm{g}_n}{\|\bm{g}_n\|^2_2} \\
		&= \left[\begin{array}{ccc}
			\frac{r_{1, n}}{r_0} \mathrm{e}^{-j k\left(r_{1, n}-r_{0}\right)}  & \ldots & \frac{r_{M, n}}{r_0} \mathrm{e}^{-j k\left(r_{M, n}-r_{0}\right)}
		\end{array}\right]^{T}.
	\end{aligned}
\end{equation}
This allows the measured multi-dimensional signals to be divided into $J$ frames to form a cross-spectral matrix (CSM), which is the correlation matrix of the signals received by the multiple channels in the frequency domain, i.e.,
\begin{equation} 
	\bm{C}=\frac{1}{J} \sum_{j=1}^{J} \bm{p}_{j}(f) \bm{p}_{j}(f)^{H},
\end{equation}
where $(\cdot)^{H}$ denotes the conjugate transpose, and
\begin{equation}
	\bm{p}_{j}(f) = \left[\begin{array}{cccc}
		p_{j,1}(f) & p_{j,2}(f) & \ldots & p_{j,M}(f)
	\end{array}\right]^{T},
\end{equation}
with $p_{j,m}(f)$ denoting the $j$-th frame frequency-domain signal of $i$-th microphone. Forming the frequency representation for all grid points, the acoustic source power, i.e., the beamforming map, may be expressed as
\begin{equation}
	\bm{b} = \left[\begin{array}{cccc}
		b_{1} & b_{2} & \ldots & b_{N^2}
	\end{array}\right]^{T},
\end{equation}
where $b_{n}$ denotes the power at the $n$-th grid point, i.e.,
\begin{equation} 
	b_n=\frac{1}{M^{2}} \bm{w}_{n}^{H} \bm{C} \bm{w}_{n}.
\end{equation}
The DAS is an efficient method for acoustic beamforming that has been widely applied in various fields, due to its robustness, and as it is computationally efficient and intuitive. However, it is subject to Sparrow's resolution limit, generally implying a low spatial resolution and the presence of spurious estimates.

\begin{figure*}[!t]
	\centering
	\includegraphics[width=\linewidth]{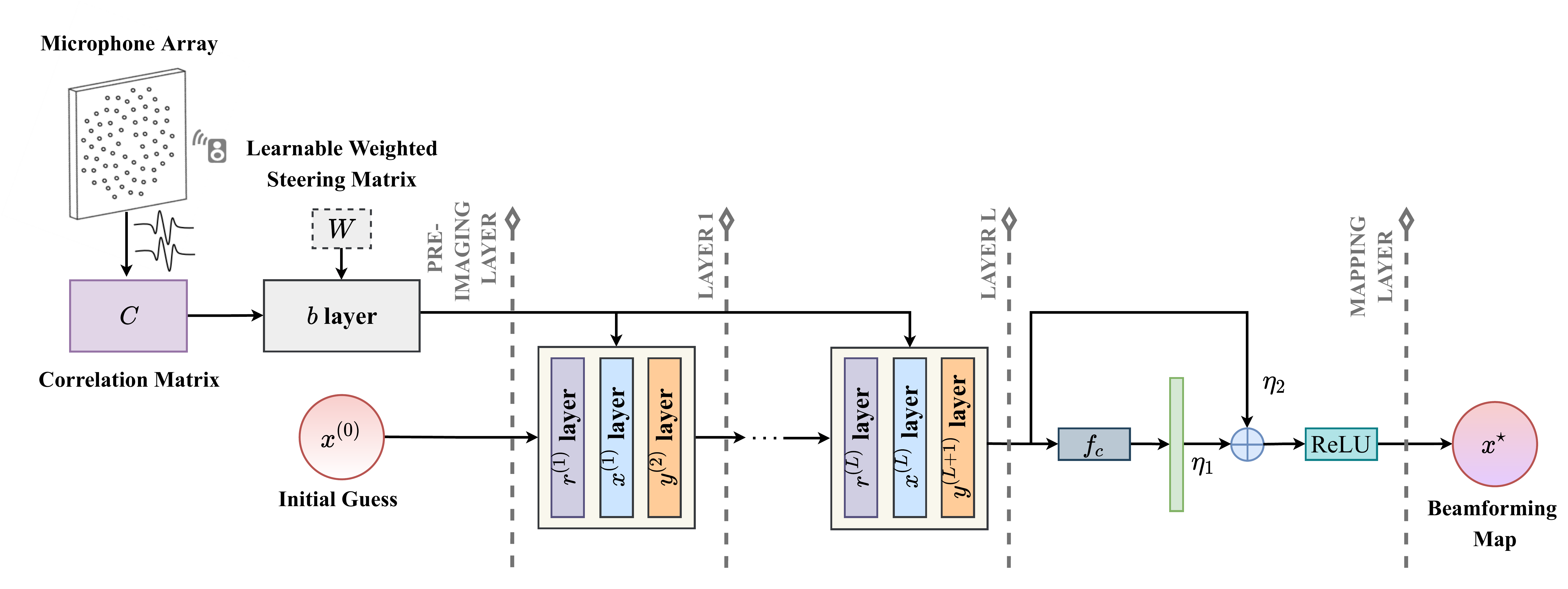}
	\caption{The overall architecture of the proposed DAMAS-FISTA-Net with $L$ iteration layers. This framework mainly consists of five types of modules: Pre-imaging ($\bm{b}$), reconstruction ($\bm{r}^{(k)}$), nonlinear transform ($\bm{x}^{(k)}$), momentum ($\bm{y}^{(k+1)}$), and mapping ($\bm{x}^*$). A correlation matrix, $\bm{C}$, directly obtained from the raw data, may be successively processed, and finally generates a beamforming map.}
	\label{fig3}
\end{figure*}

\subsection{Deconvolution Approach for Acoustic Beamforming}
Employing the characteristics of acoustic source propagation, the DAMAS \cite{brooks2006deconvolution} constructs the inverse problem
\begin{equation} \label{eq8}
	\bm{b} = \bm{A}\bm{x},
\end{equation}
where $\bm{x}$ denotes a column vector containing the unknown source power and $\bm{A}$ denotes the propagation matrix, i.e.,
\begin{equation} 
	\bm{A} = \frac{1}{M^2} \text{abs}(\bm{W}^*\bm{G}^T) \circ \text{abs}(\bm{W}^*\bm{G}^T),
\end{equation}
where $\bm{G}\in R^{N^2\times M}$ and $\bm{W}\in R^{N^2\times M}$ denote the steering matrix and the weighted steering matrix, respectively, such that
\begin{equation}
	\bm{G} = \left[\begin{array}{c}
		\bm{g}_1^T \\
		\ldots \\
		\bm{g}_{N^2}^T
	\end{array}\right], \enspace
	\bm{W} = \left[\begin{array}{c}
		\bm{w}_1^T \\
		\ldots \\
		\bm{w}_{N^2}^T
	\end{array}\right],
\end{equation}
with $\circ$ denoting the Hadamard product, $(\cdot)^*$ the conjugate, and $\text{abs}(\cdot)$ the element-wise absolute value. The resulting inverse problem in (\ref{eq8}) may be solved efficiently using, for instance, a Gauss–Seidel iterative method. We refer the reader to \cite{brooks2006deconvolution} where the details of the resulting optimization process are discussed. The overall computational complexity of the DAMAS algorithm is $O(N^6)$, making it problematic to employ in real-time acoustic beamforming.

\section{The proposed method} \label{s4}

In this section, we first propose an iterative solver, dubbed the DAMAS-FISTA, to improve the efficiency of the DAMAS estimator. We then proceed to construct the DAMAS-FISTA-Net, an interpretable end-to-end network, by studying the DAMAS-FISTA update equations.

\subsection{The DAMAS-FISTA Algorithm}

Inspired by \cite{ehrenfried2007comparison}, assuming that the DAS result, $\bm{b}$, can be modelled as experiencing a white Gaussian noise, and exploiting the prior information that the source power, $\bm{x}$, is nonnegative, we propose to recover the sound power map by solving the optimization problem
\begin{equation} \label{eq11}
	\begin{array}{ll}
		\min \limits_{\bm{x}}  \quad\quad \frac{1}{2} \|\bm{A}\bm{x}-\bm{b}\|_{2}^{2} \\
		\ \text{s.t.} \enspace \quad\quad x_i \geq 0.
	\end{array}
\end{equation}

Since (\ref{eq11}) is a non-negative convex optimization problem, the FISTA, which preserves the computational simplicity of ISTA but with a global rate of convergence which is significantly improved, both theoretically and practically \cite{daubechies2004iterative,beck2009fast}, may be introduced to solve it by iterating the following update steps:
\begin{align}
	\bm{r}^{(k)}&=\bm{y}^{(k)}-\frac{1}{\mathscr{L}} \bm{A}^{T}\left(\bm{A} \bm{y}^{(k)}-\bm{b}\right), \label{eq12}\\
	\bm{x}^{(k)}&=\max\left(\bm{r}^{(k)},0\right),\\
	t^{(k+1)} &= \frac{1+\sqrt{1+4(t^{(k)})^2}}{2},\\
	\bm{y}^{(k+1)} &=\bm{x}^{(k)} + \frac{t^{(k)}+1}{t^{(k+1)}} \left(\bm{x}^{(k)}-\bm{x}^{(k-1)}\right), \label{eq15}
\end{align}
where $\mathscr{L}$ denotes a Lipschitz constant, i.e., the maximum eigenvalue of the Hessian matrix $\bm{A}^T\bm{A}$, whereas $\bm{r}^{(k)}$, $t^{(k)}$, and $\bm{y}^{(k)}$ are auxiliary variables. The proposed DAMAS-FISTA algorithm is summarized in Algorithm \ref{alg1}. It is suggested that the stopping threshold $\epsilon$ is selected in the range $10^{-2}$ to $10^{-4}$, with the vectors $\bm{x}^{(0)}$ and $\bm{y}^{(1)}$ being initialized as a zero vector.

\begin{algorithm}[!t] 
	\caption{The DAMAS-FISTA algorithm}  
	\label{alg1}  
	\begin{algorithmic} 
		\Require
		The Lipschiztz constant, $\mathscr{L}$, the initial guess, $\bm{x}^{(0)}$, the auxiliary vector, $\bm{y}^{(1)}$, the step parameter, $\bm{t}^{(1)} = 1$, the stopping threshold, $\varepsilon$, and the maximum number of iterations. 
		\Repeat  
		\State step1. Update $\bm{x}^{(k)}$
		$$\bm{x}^{(k)}=\max\left(\bm{y}^{(k)}-\frac{1}{\mathscr{L}} \bm{A}^{T}\left(\bm{A} \bm{y}^{(k)}-\bm{b}\right),0\right),$$
		\State step2. Update $t^{(k+1)}$  
		$$t^{(k+1)} = \frac{1+\sqrt{1+4(t^{(k)})^2}}{2},$$
		\State step3. Update $\bm{y}^{(k+1)}$
		$$\bm{y}^{(k+1)} =\bm{x}^{(k)} + \frac{t^{(k)}+1}{t^{(k+1)}} \left(\bm{x}^{(k)}-\bm{x}^{(k-1)}\right),$$ 
		\Until $\|\bm{x}^{(k)}-\bm{x}^{(k-1)}\|_{2} \textless \varepsilon \| \bm{x}^{(k-1)}\|_{2}$ or $k$ reaches the maximum number of iterations.
		\Ensure The acoustic beamforming result, $\bm{x}^{*}$.
	\end{algorithmic}  
\end{algorithm}

\subsection{Network Architecture}
It is worth noting that the above DAMAS-FISTA estimator is an iterative solver, for which one may, reminiscent to \cite{gregor2010learning,wang2016learning,DBLP:conf/nips/Yang0LX16,NEURIPS2019_e9bf14a4,yang2018admm,zhang2018ista,xiang2021fista}, construct a corresponding interpretable end-to-end network architecture, here termed the DAMAS-FISTA-Net. The basic idea in order to do so is to design a non-linear, parameterized, feedforward architecture with a fixed depth, $L$, which can be trained to approximate the optimal beamforming map. Notably, in contrast to the existing networks for acoustic beamforming, we here design a steering matrix learning module, which can directly construct a beamforming map from raw data, forming a more flexible end-to-end network. DAMAS-FISTA-Net is tailored to the acoustic beamforming problem, mainly generalizing the five types of operations to have learnable parameters as network layers, i.e., a pre-imaging layer ($\bm{b}$), reconstruction layers ($\bm{r}^{(k)}$), nonlinear transform layers ($\bm{x}^{(k)}$), momentum layers ($\bm{y}^{(k+1)}$), and a mapping layer ($\bm{x}^{*}$). Among them, the $\bm{r}^{(k)}$ layers, the $\bm{x}^{(k)}$ layers, and the $\bm{y}^{(k+1)}$ layers are collectively termed the iteration layers forming a fixed depth, $L$. Fig.\ref{fig3} illustrates the architecture of the proposed DAMAS-FISTA-Net. We proceed to examine these layers in further detail.

\textbf{Pre-imaging layer $\bm{b}$:}
This layer directly processes the raw data (the CSM of the input signal), and then generates a pre-imaging power map, $\bm{b}$, corresponding to the DAS procedure. As mentioned above, there are several options for the weighted steering matrix. Here, we allow this matrix to be learned directly from the sound field information in a data-driven manner. Specifically, given the correlation matrix, $\bm{C}$, the output of this layer may be written as
\begin{equation}
	\bm{b} = \frac{1}{M^2} \text{Real}\left(\text{RowSum}\left(\bm{W}^* \circ \left(\bm{C} \cdot \bm{W}\right) \right)^T\right),
\end{equation}
where $\text{Real}(\cdot)$ denotes the operation that returns the real part, and $\text{RowSum}(\cdot)$ the operation that returns the sum of each row of the input matrix, with $\bm{W}$ representing the learnable weighted steering matrix.

\textbf{Reconstruction layer $\bm{r}^{(k)}$:}
This layer reconstructs the power map based on the gradient descent operation of (\ref{eq12}). In order to reduce parameter redundancy, the reconstructed dictionary, $\bm{A}$, is fixed. Simultaneously, to preserve the DAMAS-FISTA structure while increasing the network flexibility, the step size may be allowed to vary across iterations, and a weighted parameter may be used. Then, given $\bm{y}^{(k)}$, the output of this layer is defined as
\begin{equation}
	\bm{r}^{(k)}=\iota^{(k)}\bm{y}^{(k)}-\rho^{(k)} \bm{A}^{T}\left(\bm{A} \bm{y}^{(k)}-\bm{b}\right).
\end{equation}

\textbf{Nonlinear transform layer $\bm{x}^{(k)}$:}
This layer aims to ensure the non-negativity of the power estimate by limiting the reconstructed result to lie in a non-negative domain. Given $\bm{r}^{(k)}$, i.e., the reconstructed result in stage $k$, the output is
\begin{equation}
	\bm{x}^{(k)}=\text{ReLU}\left(\bm{r}^{(k)}\right),
\end{equation}
where $\text{ReLU}(\cdot)$ denotes the linear rectification function, i.e., 
\begin{equation}
	\text{ReLU}(\bm{x})=\max(\bm{x},0).
\end{equation}

\textbf{Momentum layer $\bm{y}^{(k+1)}$:}
This layer performs a momentum update operation by replacing the constant update step, $t^{(k)}$, in (\ref{eq15}) with two learnable parameters, $\tau^{(k)}$ and $\mu^{(k)}$, making the network learn flexible features autonomously from dataset. Given $\bm{x}^{k-1}$ and $\bm{x}^{(k)}$, the output may be expressed as
\begin{equation}
	\bm{y}^{(k+1)} = \tau^{(k)} \bm{x}^{(k)} + \mu^{(k)} \left(\bm{x}^{(k)}-\bm{x}^{(k-1)}\right).
\end{equation}

\textbf{Mapping layer $\bm{x}^{*}$:}
In this layer, inspired by \cite{he2016deep}, a weighted residual learning module may be designed to map the preliminary output of the $L$-th nonlinear transform layer to the target domain. Specifically, we utilize a fully-connected layer, $f_c$, to learn a residual function with reference to the layer inputs. Then, a weighted attention mechanism may be used to learn the mapping feature. Notably, the output power should be non-negative. Thus, the output of this layer with input $\bm{x}^{(L)}$ is finally defined as
\begin{equation}
	\bm{x}^{*} = \text{ReLU}\left(\eta_1 \bm{x}^{(L)}+\eta_2 f_c(x^{(L)})\right),
\end{equation}
where $\eta_1$ and $\eta_2$ are both learnable weighted parameters, whereas $\bm{x}^{*}$ denote the acoustic beamforming result learned by the DAMAS-FISTA-Net.

\subsection{Network Parameters}
The learnable parameters of the DAMAS-FISTA-Net are
\begin{equation}
	\Theta = \{\bm{W}\} \cup\{\iota^{(k)}, \rho^{(k)}, \tau^{(k)}, \mu^{(k)}\}_{k=1}^L \cup \{\eta_1, \eta_2, f_c(\theta)\},
\end{equation}
where $L$ denotes the depth of the iteration layers and $f_c(\theta)$ the parameters in the fully-connected layer. All these parameters are considered as the network parameters to be learned.

\subsection{Initialization}
Instead of random initialization, we propose to employ a warm-starting manner to initialize the network. Specifically, the weighted steering vector, $\bm{W}^{(0)}$, may be initialized according to (\ref{eq3}), whereas the weighted parameter, $\rho^{(k)}$, can be initialized as the maximum eigenvalue of $\bm{A}^T \bm{A}$. For the initial guess, $\bm{x}^{(0)}$ may be initialized as a zero vector, as well as $\bm{y}^{(1)}=\bm{x}^{(0)}$. The fully-connected layer, $f_c$, is initialized with Xavier's algorithm \cite{pmlr-v9-glorot10a}. The parameters $\{\iota^{(k)}, \tau^{(k)}, \mu^{(k)}\}_{k=1}^L$ and $\{\eta_1, \eta_2\}$ are all initialized as 1.

\subsection{Loss Function Design}
Given a training data pairs ($\bm{p}_i$, $\bm{b}^{(gt)}_i$), the DAMAS-FISTA-Net first transforms the raw data from microphone array, $\bm{p}_i$, to a cross-spectral matrix without information loss. Then, the cross-spectral matrix may be taken as input and generates a high-quality beamforming map, $\hat{\bm{b}}_i(\bm{p}_i,\Theta)$, as output. A training set $\Gamma$ is constructed containing pairs of raw data and ground-truth. The network parameters are optimized by minimizing the mean squared error (MSE) loss function, i.e.,
\begin{equation}
	\mathcal{L}_{mse} = \frac{1}{|\Gamma|} \! \sum_{(\bm{p}_i, \bm{b}^{(gt)}_i)\in \Gamma} \! \sqrt{\|\hat{\bm{b}}_i(\bm{p}_i,\Theta) - \bm{b}^{(gt)}_i\|_2^2},
\end{equation}
which seeks to reduce the discrepancy between the network estimate and the ground truth.

\subsection{Implementation Details}

In our work, we use the PyTorch \cite{paszke2017automatic} framework to carry out our experiments with a single 3090Ti GPU. We employ the Adam \cite{kingma2014adam} optimizer, with the initial learning rate being $1\times10^{-3}$, the weight decay being $1\times10^{-2}$, and the momentum being 0.9. The mini-batch size during training is set to 64, and the depth of the iteration layer, $L$, is set to 5.

\section{Numerical Results and Discussions} \label{s5}

To evaluate the capability of the DAMAS-FISTA and the DAMAS-FISTA-Net\footnote{Our implementation will be available online upon publication.}, we examine both simulated and real-world data, comparing the bias, R\'enyi entropy, and runtime performance of the proposed network to alternative state-of-the-art methods, namely the model-based DAS \cite{van1988beamforming}, DAMAS \cite{brooks2006deconvolution}, FFT-NNLS \cite{ehrenfried2007comparison}, and FFT-DFISTA \cite{ding2022high} estimators, as well as the network-based methods Acoustic-AlexNet \cite{reiter2017machine},  Acoustic-ResNet \cite{kujawski2019deep}, and Acoustic-Net \cite{zhou2022acoustic}. In comparing the methods, we examine:

\begin{enumerate}
	\item[i)] The concentration of the beamforming map energy, evaluated using the R\'enyi entropy \cite{stankovic2001measure,ding2022high}, i.e.,
	\begin{equation}
		R(\alpha)=\frac{1}{1-\alpha} \log _{2}\left(\frac{\iint_{\mathbb{R}^{2}}|\bm{B}|^{\alpha} d x d y}{\iint_{\mathbb{R}^{2}}|\bm{B}| dxdy}\right),
	\end{equation}
	where $\bm{B}$ denotes the estimated beamforming map with $\alpha$ usually being chosen as 3 for evaluation.

	\item[ii)] The location bias, which forms a measure of the performance of the precision of the estimated location, being defined as
	\begin{equation}
		\Delta L = \|L_{estimated}-L_{gt}\|_2,
	\end{equation}
	where ${L}_{gt}$ and ${L}_{estimated}$ denote the true and estimated locations, respectively.
	
\end{enumerate}

\subsection{Simulated Results}

To validate these algorithms, we initially generate two types of signal sources, including both one-point and two-point situations\footnote{The simulated data will be available online upon publication.}.

\subsubsection{One-point simulation source setting} 

\begin{figure}[!t]
	\centering
	\includegraphics[width=2.5in]{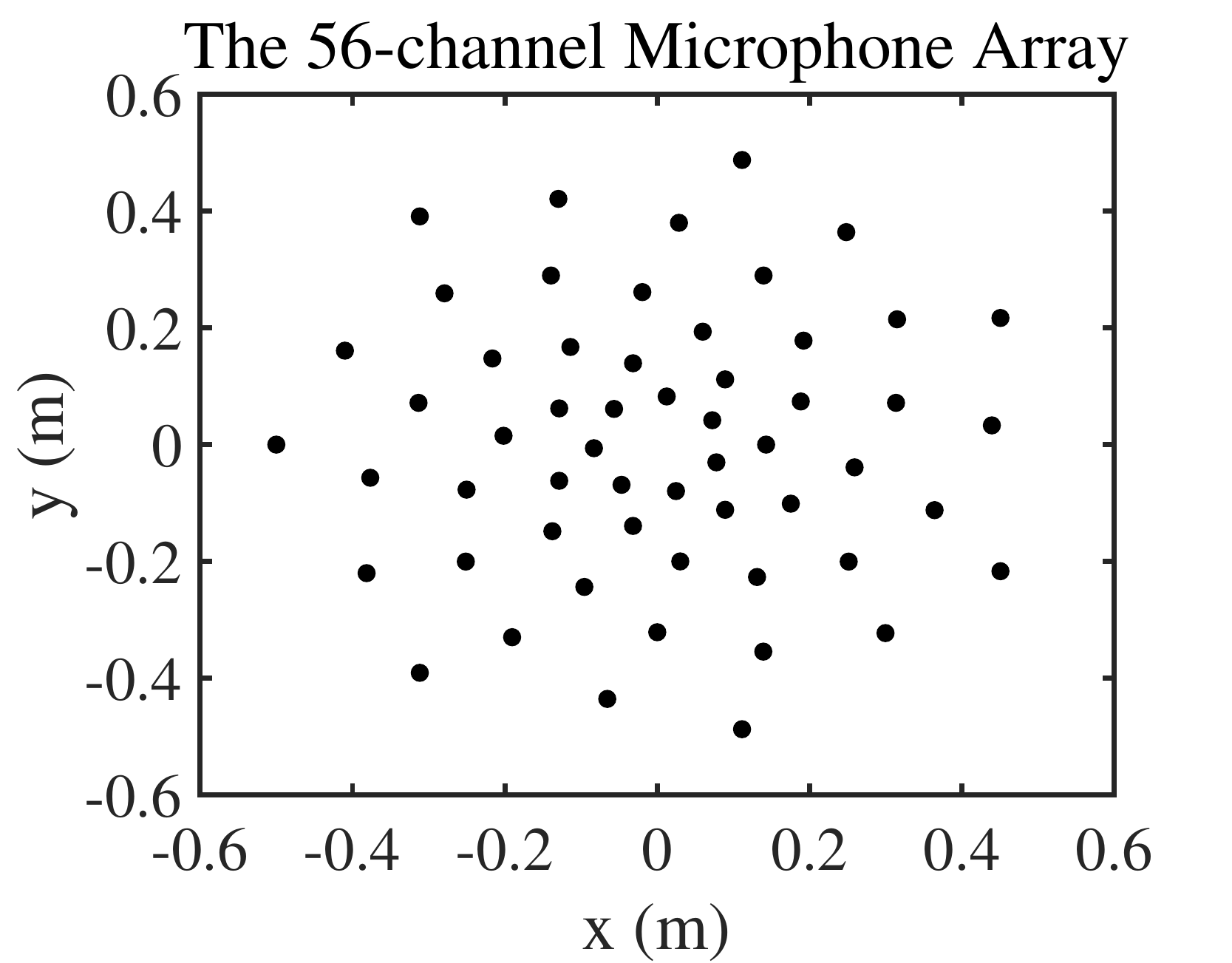}
	\caption{Spatial position of the 56-channel microphone array.}
	\label{fig4}
\end{figure}

In the one-point simulation, the signal source is formed as a sinusoidal with frequency 2000 Hz, being generated by the acoustic software Acoular \cite{sarradj22acoular}, an open source Python library, by using the 56 spiral array microphone array shown in Fig.\ref{fig4}. The signal is measured with a sampling frequency of 51200 Hz, with a sampling period of 0.02 s. Each measurement contains a single source placed in the x-y plane at a distance of $z=2.5$ m from the microphone array, with the individual source locations being sampled from a bivariate uniform distribution to form a suitable dataset of 2000 data points. The DAMAS-FISTA-Net is trained by splitting the data points into a training and validation set containing 70\% and 30\% of the data, respectively.

\subsubsection{Two-point simulation source setting} 

Each measurement results from two signal sources placed in the x-y plane at a distance of $z=2.5$ m from the microphone array, using the same signal models as in the one-point case. As before, the individual source locations are sampled from a bivariate uniform distribution to form a suitable dataset of 2000 data points. Also in this case, the DAMAS-FISTA-Net is trained by splitting the data points into a training and validation set containing 70\% and 30\% of the data, respectively.

\begin{table}[!t]
	\centering
	\caption{The experimental indicators by different algorithms for the simulated one-point case}
	\resizebox{0.5\textwidth}{!}{
	\begin{tabular}{cccccc}
		\toprule
		Methods          & $R(\alpha)$ $\downarrow$&  $\Delta L$ $\downarrow$&  Time $\downarrow$\\ \midrule
		DAS    \cite{van1988beamforming}          &     -0.2984         &              0                &      0.0655 s        \\ 
		DAMAS     \cite{brooks2006deconvolution}       &      -6.6220                      &       0        &       11.9595 s         \\ 
		FFT-NNLS     \cite{ehrenfried2007comparison}       &     -2.2609                       &   0.0987                        &  0.3547 s          \\ 
		FFT-DFISTA    \cite{ding2022high}     &     -4.1067                         &        0.1151                    &   5.2308 s          \\ \midrule
		Acoustic-AlexNet \cite{reiter2017machine} & -                               &             0               &   0.0721 s          \\ 
		Acoustic-ResNet  \cite{kujawski2019deep} &  -                            &     0.0288                       &   0.0711 s          \\ 
		Acoustic-Net   \cite{zhou2022acoustic}  & -                           &           0.0231               &  0.0588 s        \\ \midrule
		DAMAS-FISTA  (Ours)    &      \textbf{-6.6431}                     &              0          &        1.0689 s        \\ 
		DAMAS-FISTA-Net  (Ours) &      -6.5235                       &       0.0028           &         \textbf{0.0125 s}        \\ \bottomrule
	\end{tabular}
	}
	\label{table1}
\end{table}

\begin{table}[!t]
	\centering
	\caption{The experimental indicators by different algorithms for the simulated two-point case}
	\resizebox{0.5\textwidth}{!}{
		\begin{tabular}{cccccc}
			\toprule
			Methods          & $R(\alpha)$ $\downarrow$&  $\Delta L$ $\downarrow$& Time $\downarrow$\\ \midrule
			DAS    \cite{van1988beamforming}          &         0.8298     &    \textbf{0.0174}     & 0.0654 s       \\ 
			DAMAS     \cite{brooks2006deconvolution}       &          -4.4065                  &    0.0371                        &   8.4904 s         \\ 
			FFT-NNLS     \cite{ehrenfried2007comparison}       &         -1.1762                  &     0.1703                       &   0.5524 s        \\ 
			FFT-DFISTA    \cite{ding2022high}     &        -2.8402                     &      0.1639  &     6.3166 s         \\ \midrule
			DAMAS-FISTA  (Ours)    &           -4.6407               &        0.0314                 &     1.4553 s        \\ 
			DAMAS-FISTA-Net  (Ours) &         \textbf{-5.5567}                    &        0.0584                 &      \textbf{0.0124 s}      \\ \bottomrule
		\end{tabular}
	}
	\label{table2}
\end{table}

\begin{figure*}
	\centering
	\subfigure[]{\includegraphics[width=2in]{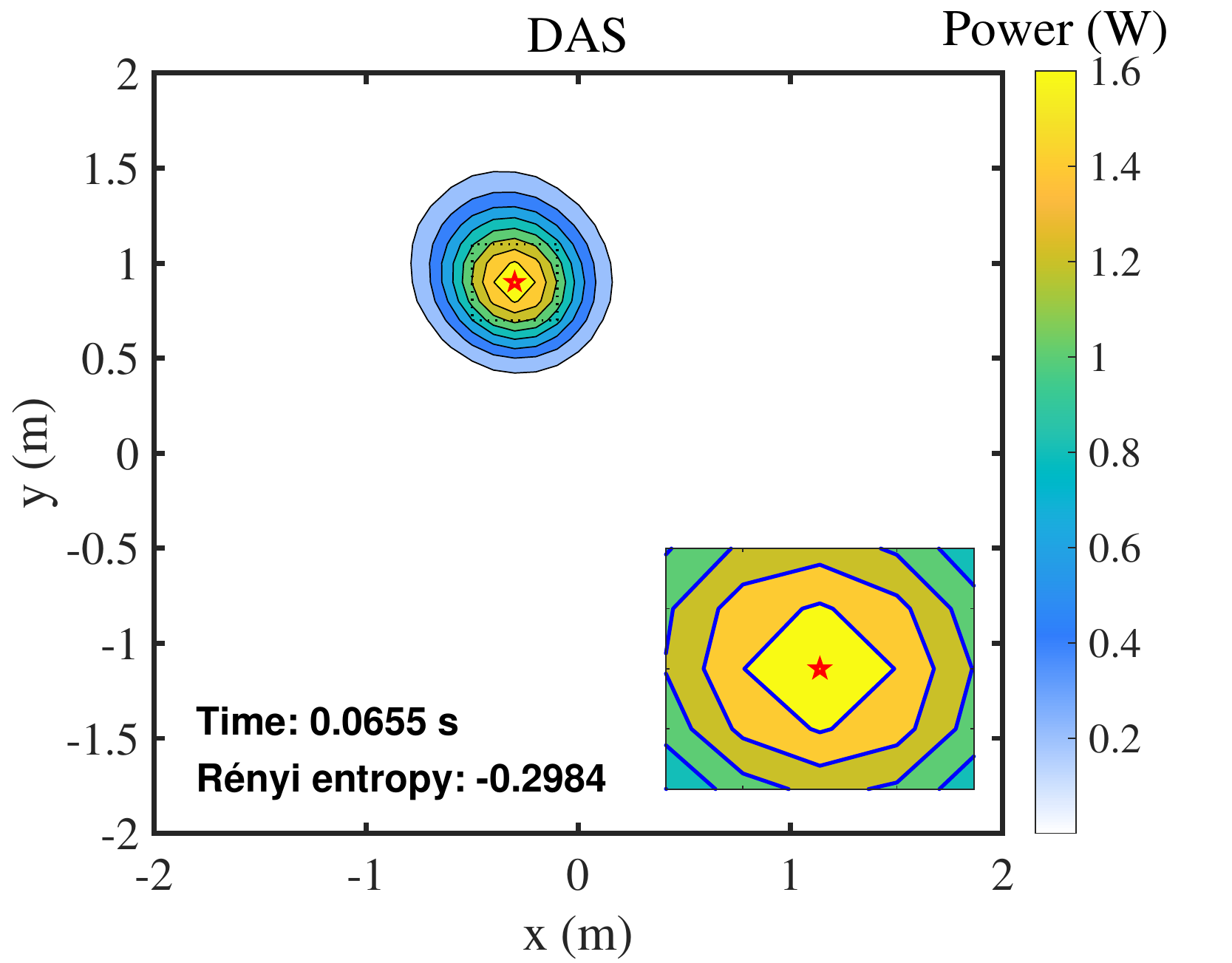}
	}
	\subfigure[]{\includegraphics[width=2in]{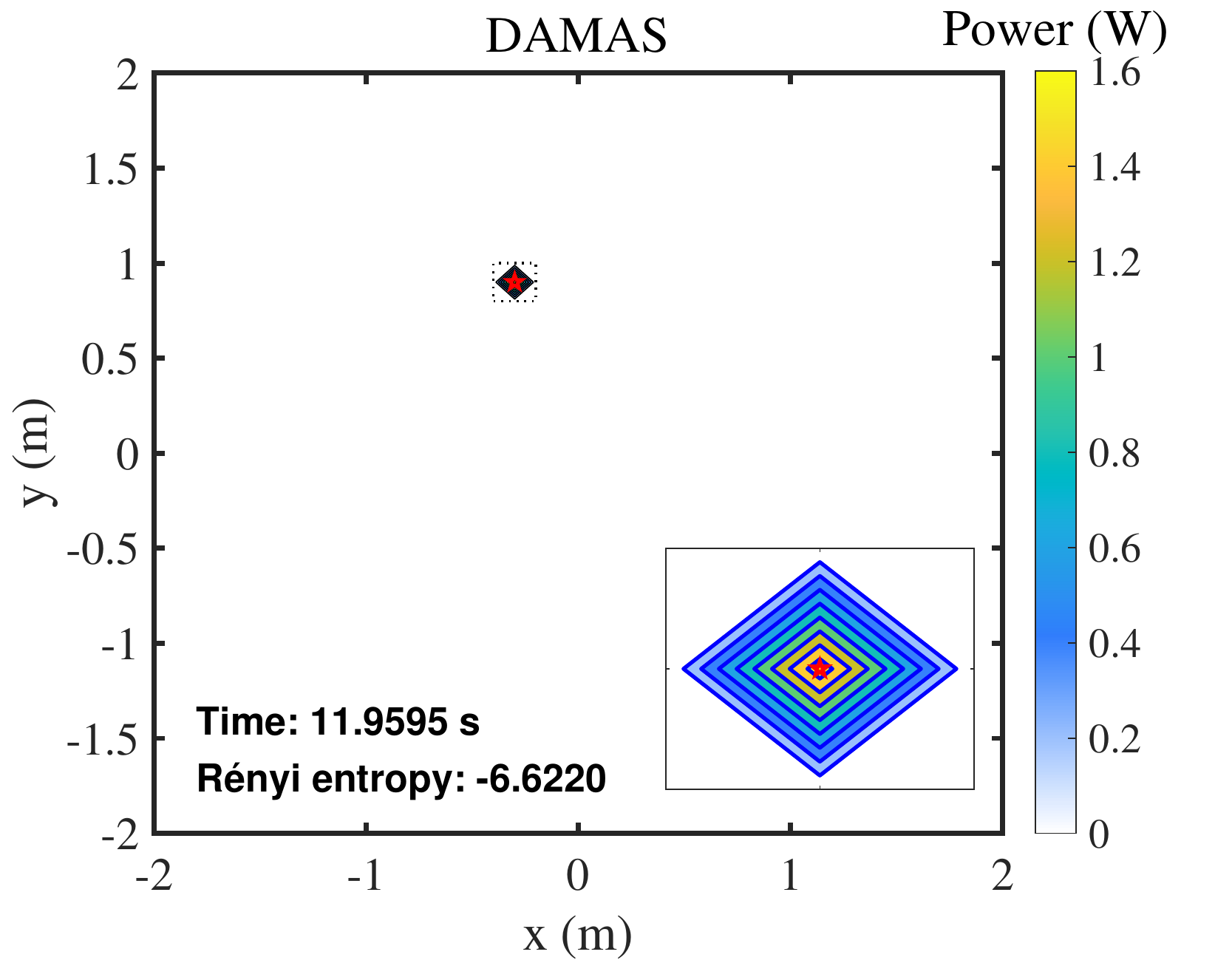}
	}
	\subfigure[]{\includegraphics[width=2in]{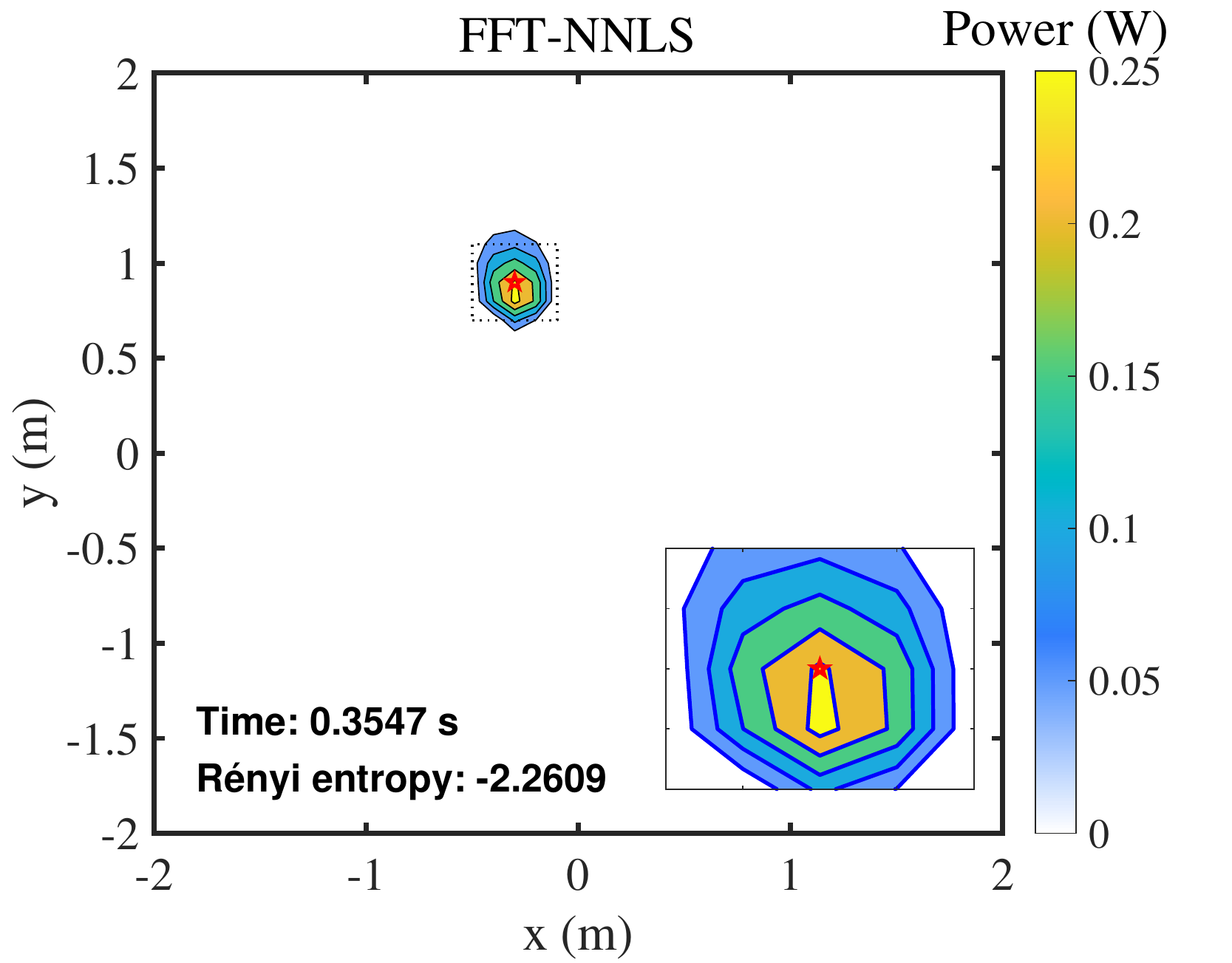}
	}
	\hfil\hfil\\
	\subfigure[]{\includegraphics[width=2in]{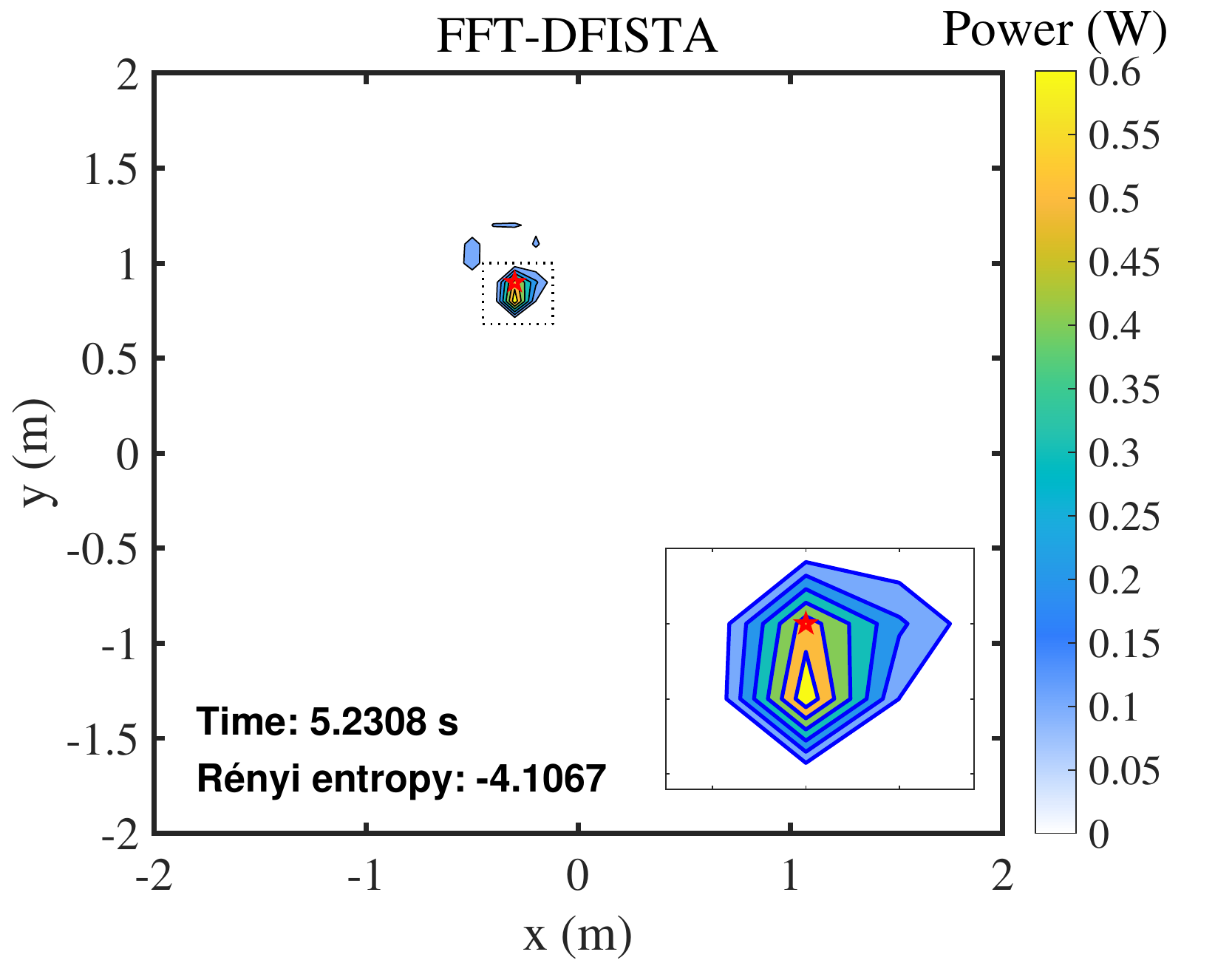}
	}
	\subfigure[]{\includegraphics[width=2in]{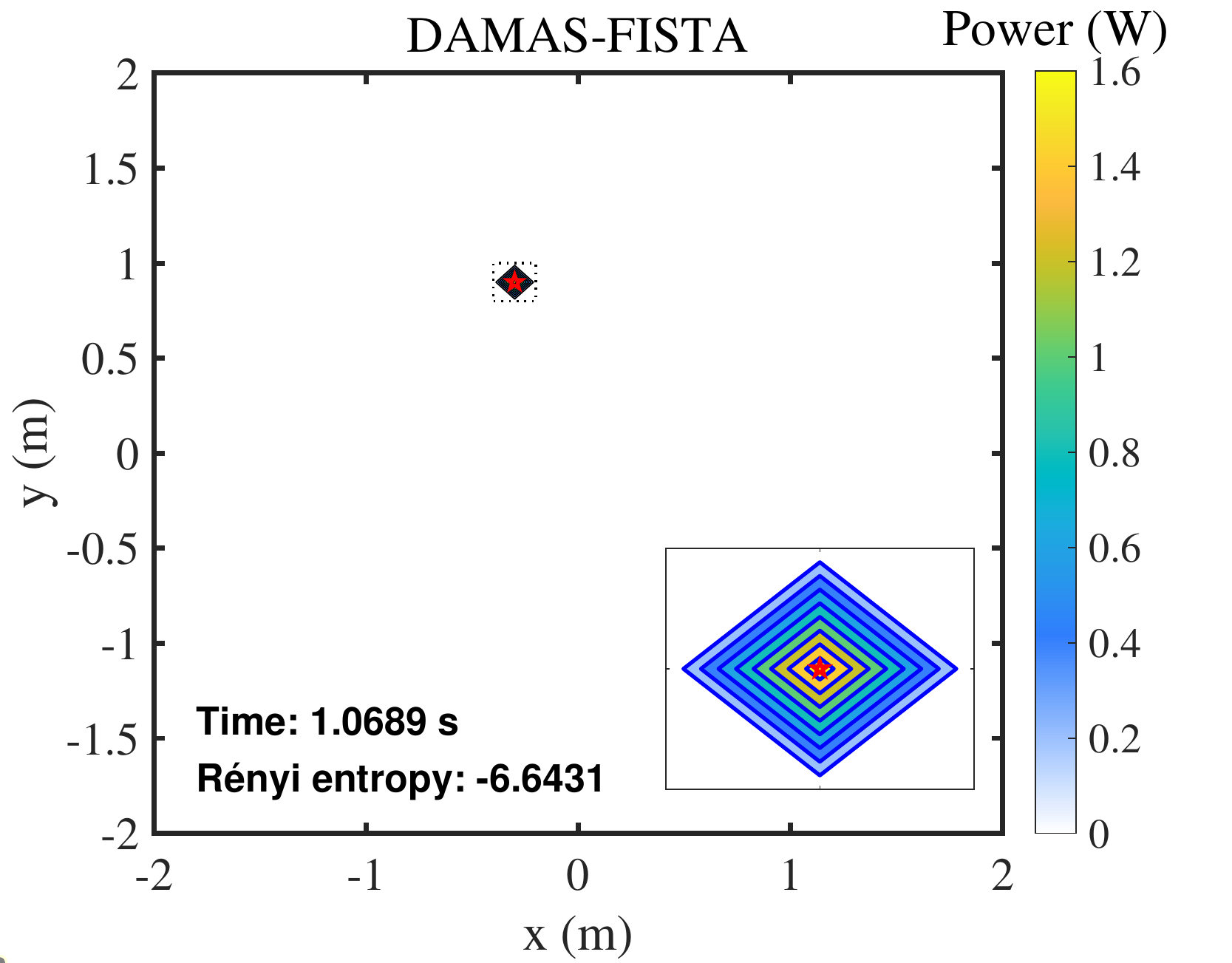}
	}
	\subfigure[]{\includegraphics[width=2in]{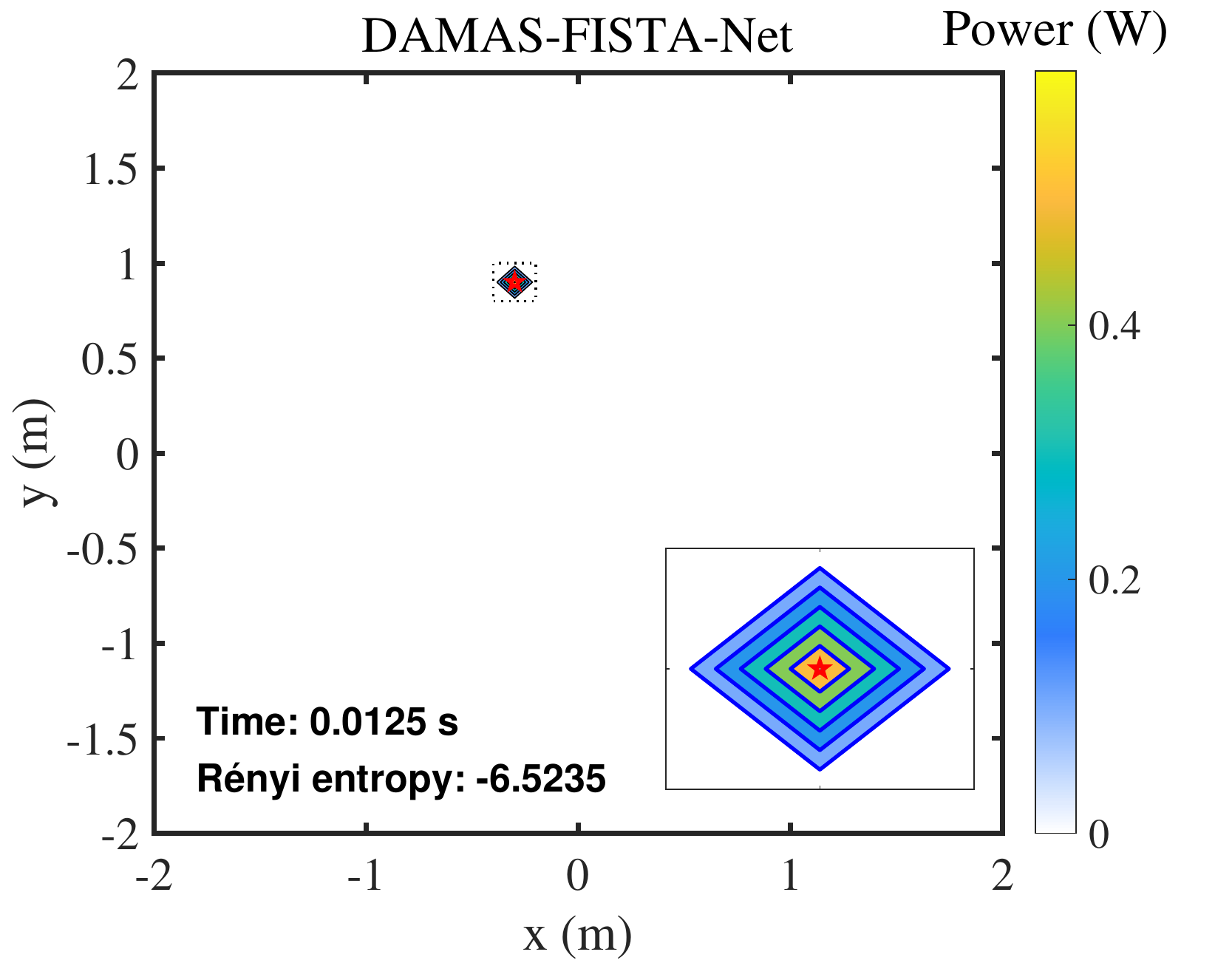}
	}
	\caption{The beamforming power map of a simulated single-point source by different algorithms (the red "star" is the true location of the source): (a) DAS, (b) DAMAS, (c) FFT-NNLS, (d) FFT-DFISTA, (e) DAMAS-FISTA, and (f) DAMAS-FISTA-Net.}
	\label{fig5}
\end{figure*}

\begin{figure*}
	\centering
	\subfigure[]{\includegraphics[width=2in]{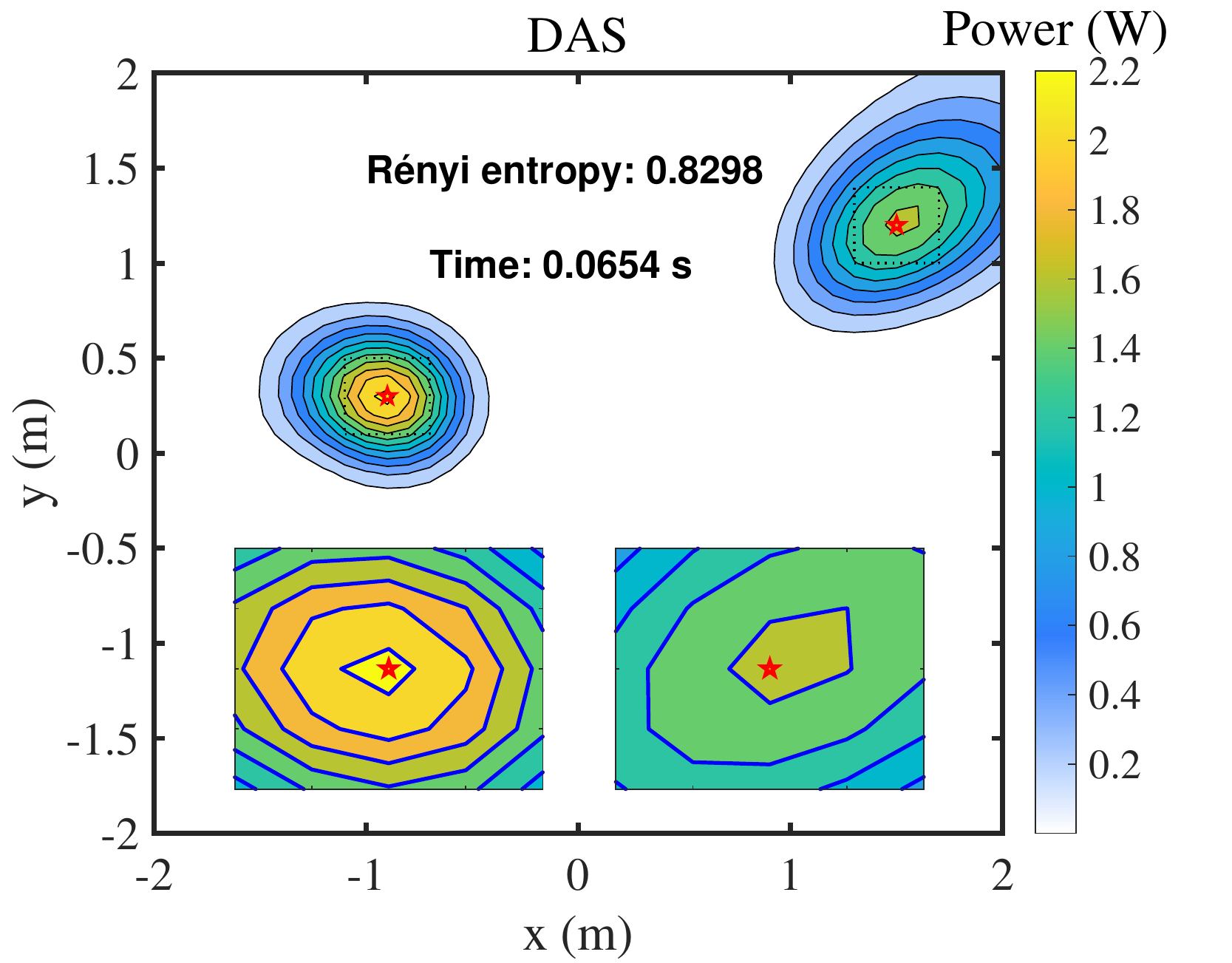}
	}
	\subfigure[]{\includegraphics[width=2in]{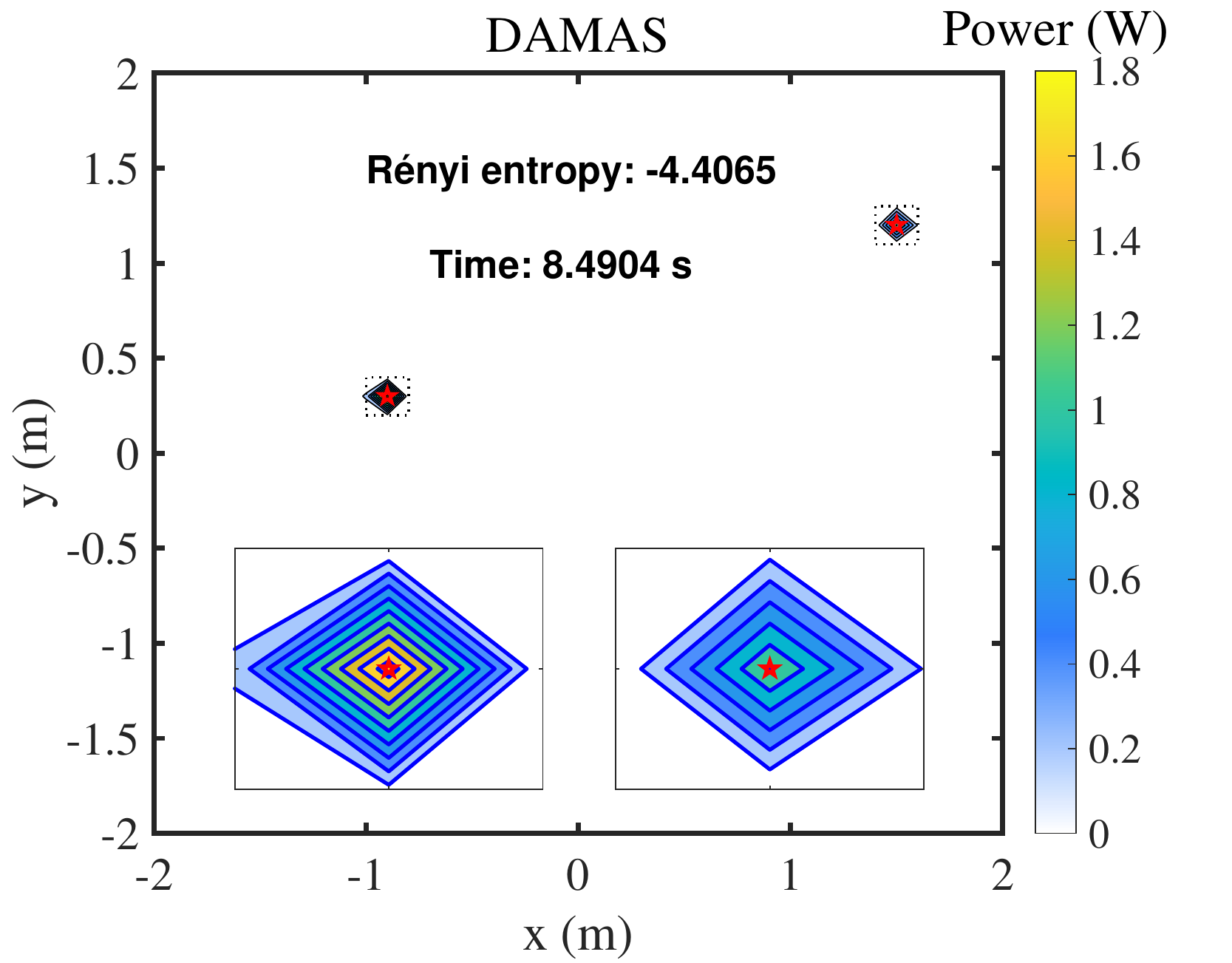}
	}
	\subfigure[]{\includegraphics[width=2in]{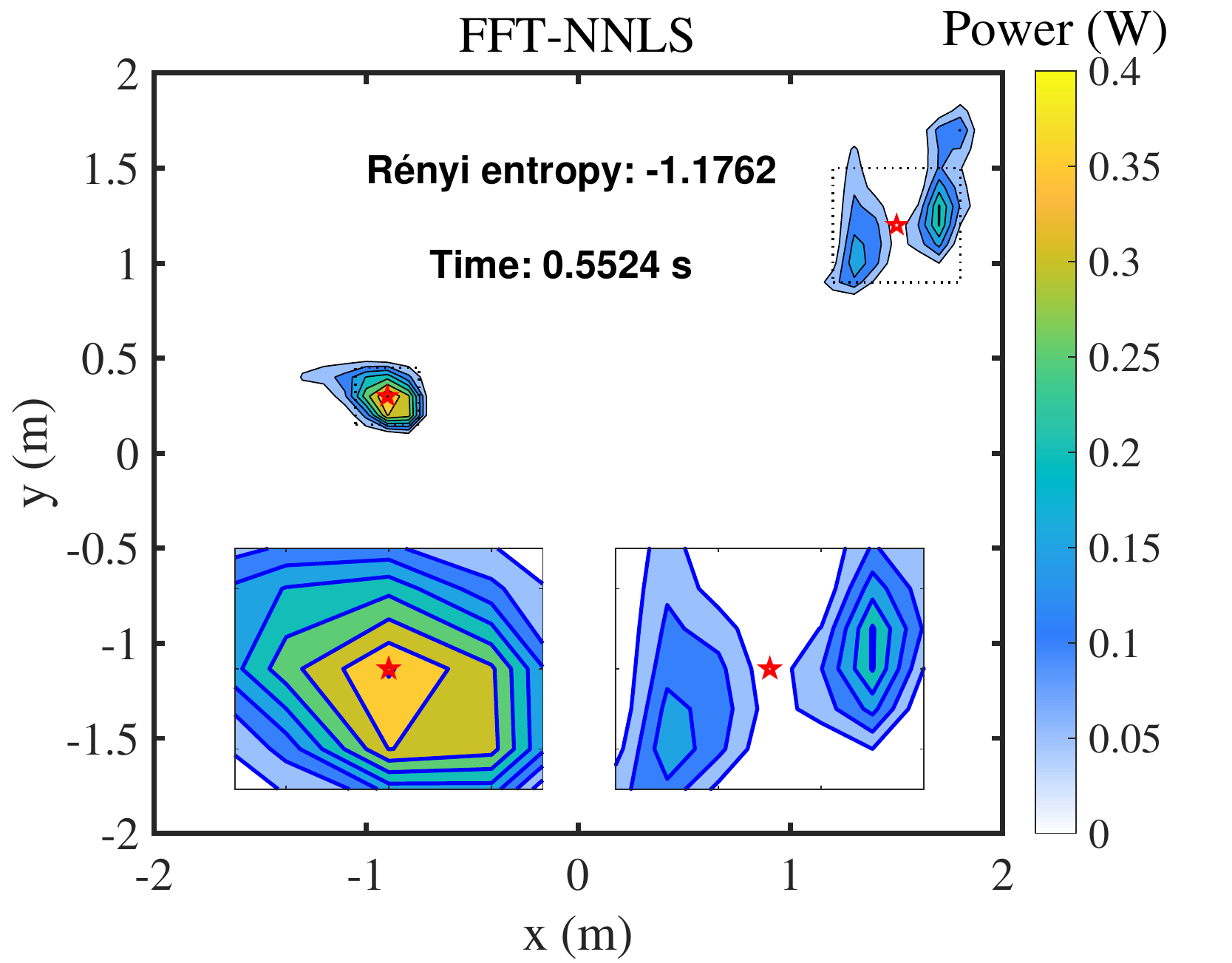}
	}
	\hfil\hfil\\
	\subfigure[]{\includegraphics[width=2in]{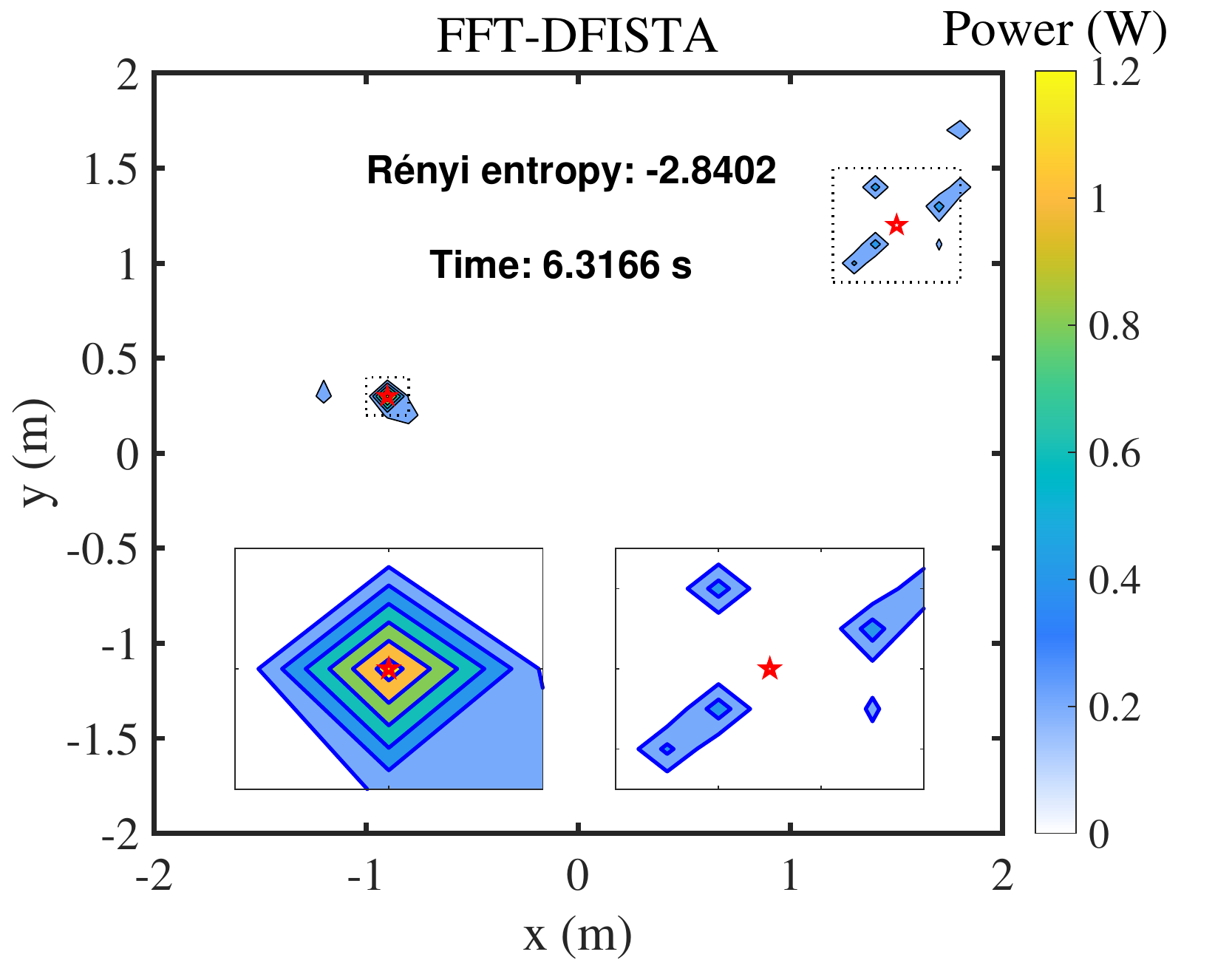}
	}
	\subfigure[]{\includegraphics[width=2in]{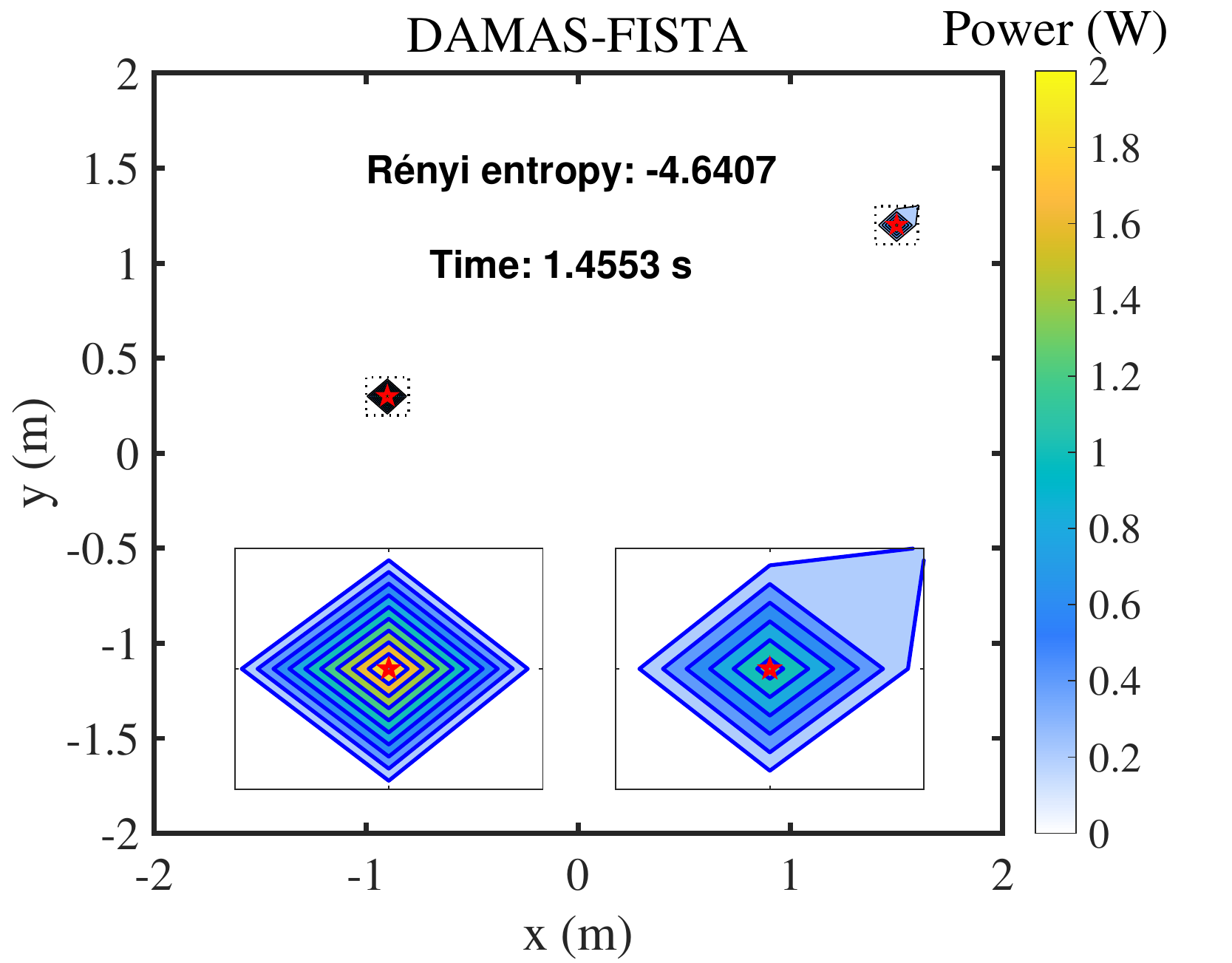}
	}
	\subfigure[]{\includegraphics[width=2in]{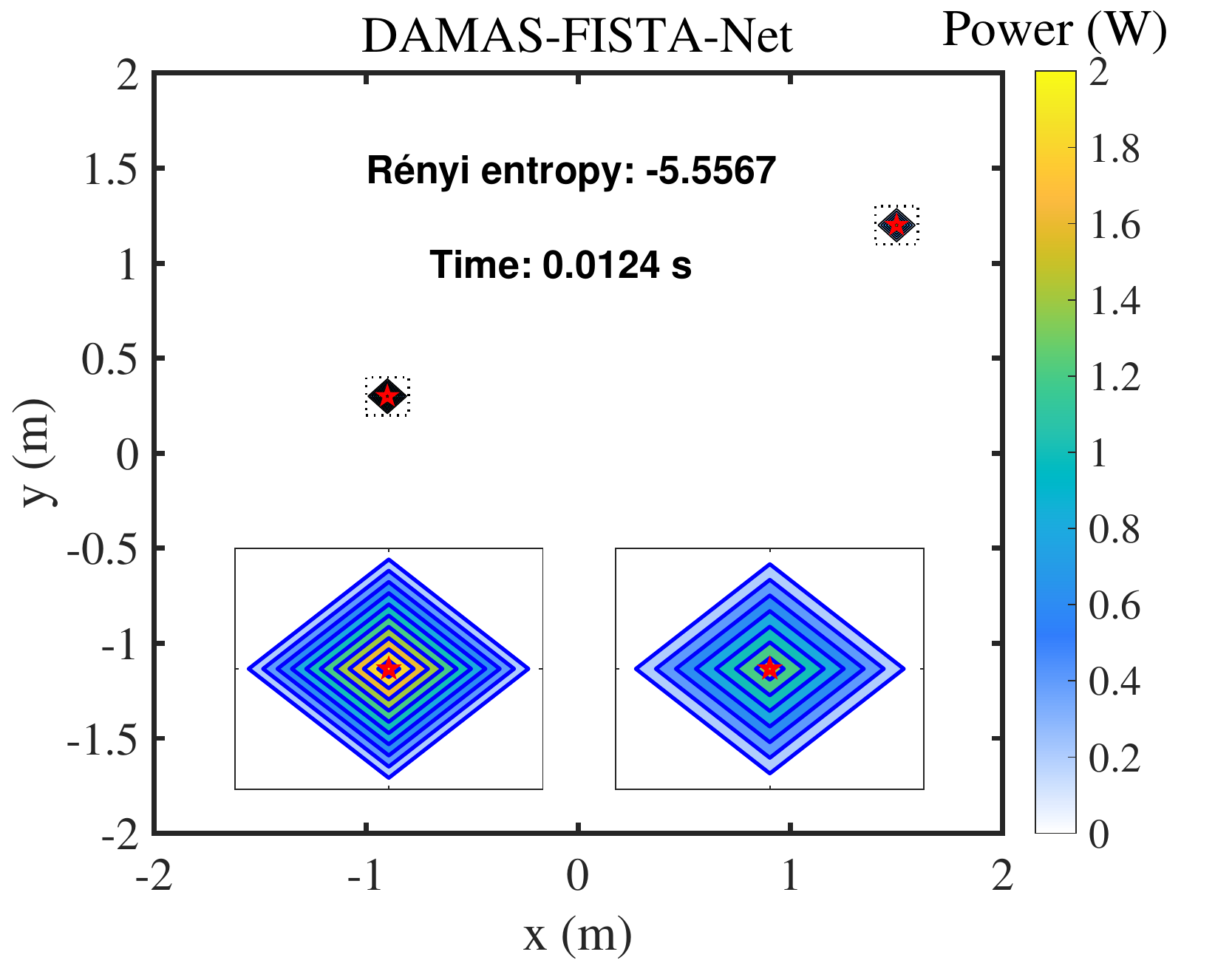}
	}
	\caption{The beamforming power map of a simulated two-point source by different algorithms (the red "star" is the true location of the source): (a) DAS, (b) DAMAS, (c) FFT-NNLS, (d) FFT-DFISTA, (e) DAMAS-FISTA, and (f) DAMAS-FISTA-Net.}
	\label{fig6}
\end{figure*}

\begin{figure*}
	\centering
	\subfigure[]{\includegraphics[width=2.3in]{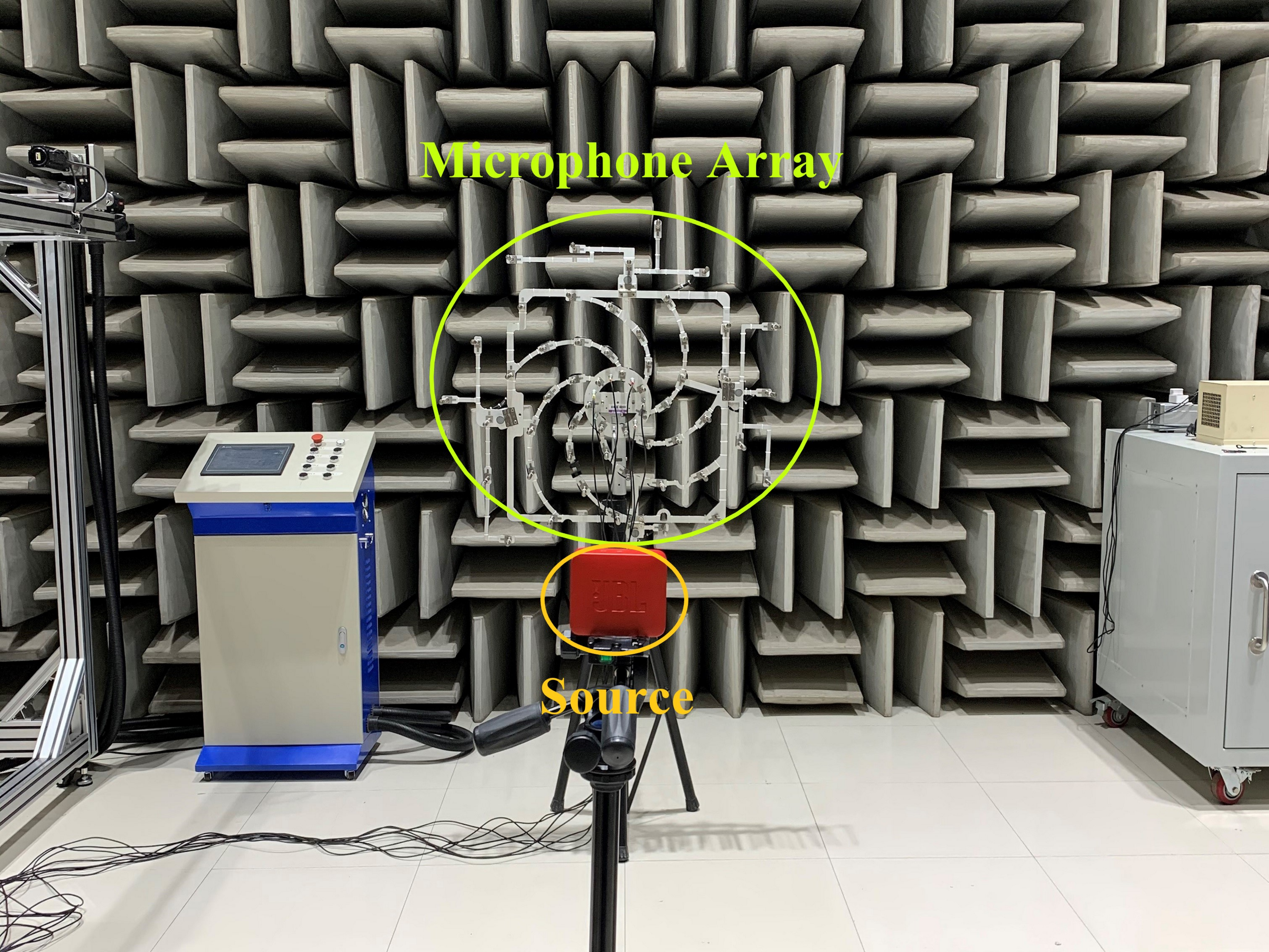}
	}
	\
	\subfigure[]{\includegraphics[width=2.3in]{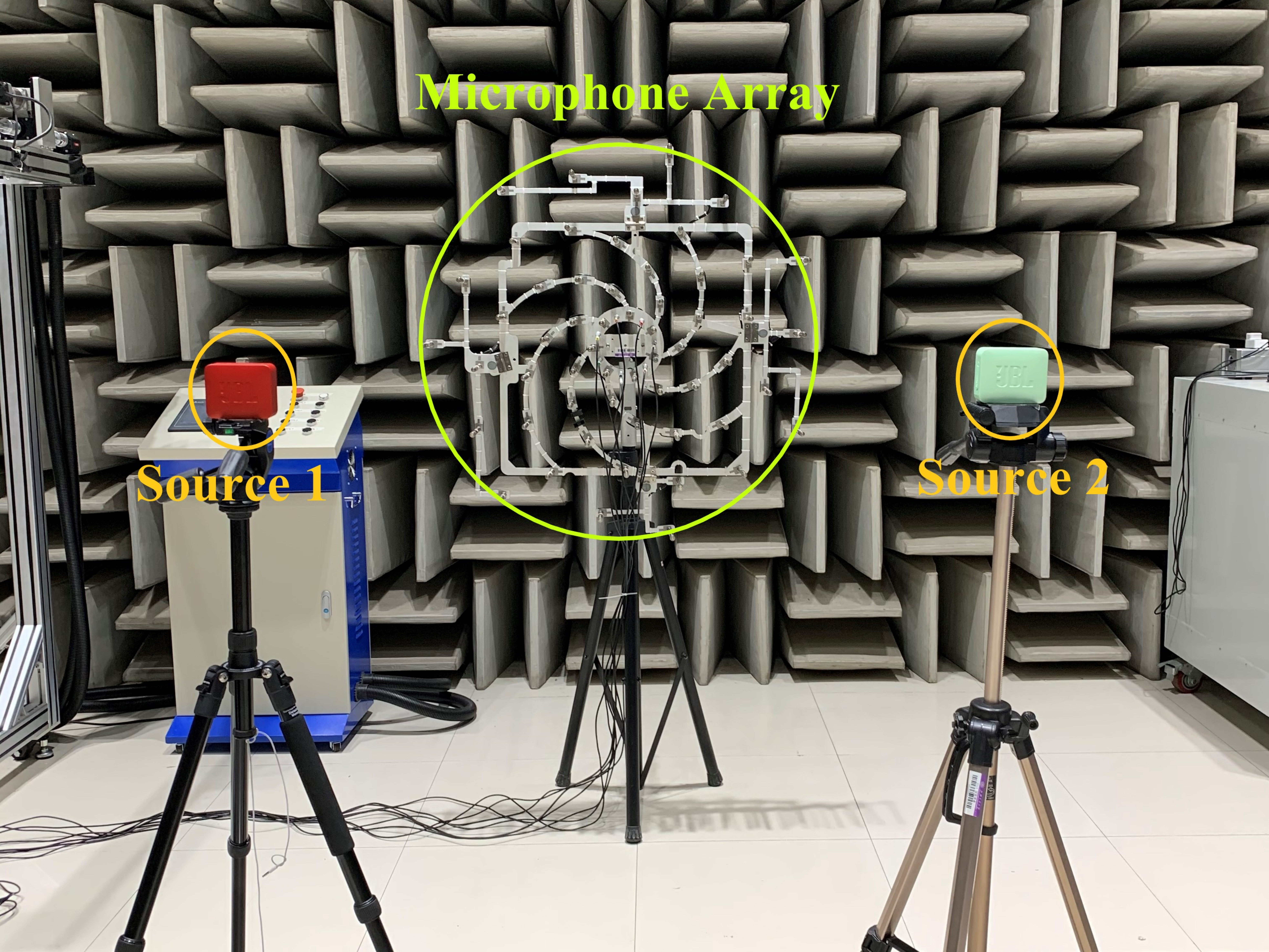}
	}
	\caption{The layout experimental setup: (a) a one-point source, (b) a two-point source.}
	\label{fig7}
\end{figure*}

\begin{figure*}
	\centering
	\subfigure[]{\includegraphics[width=1.6in]{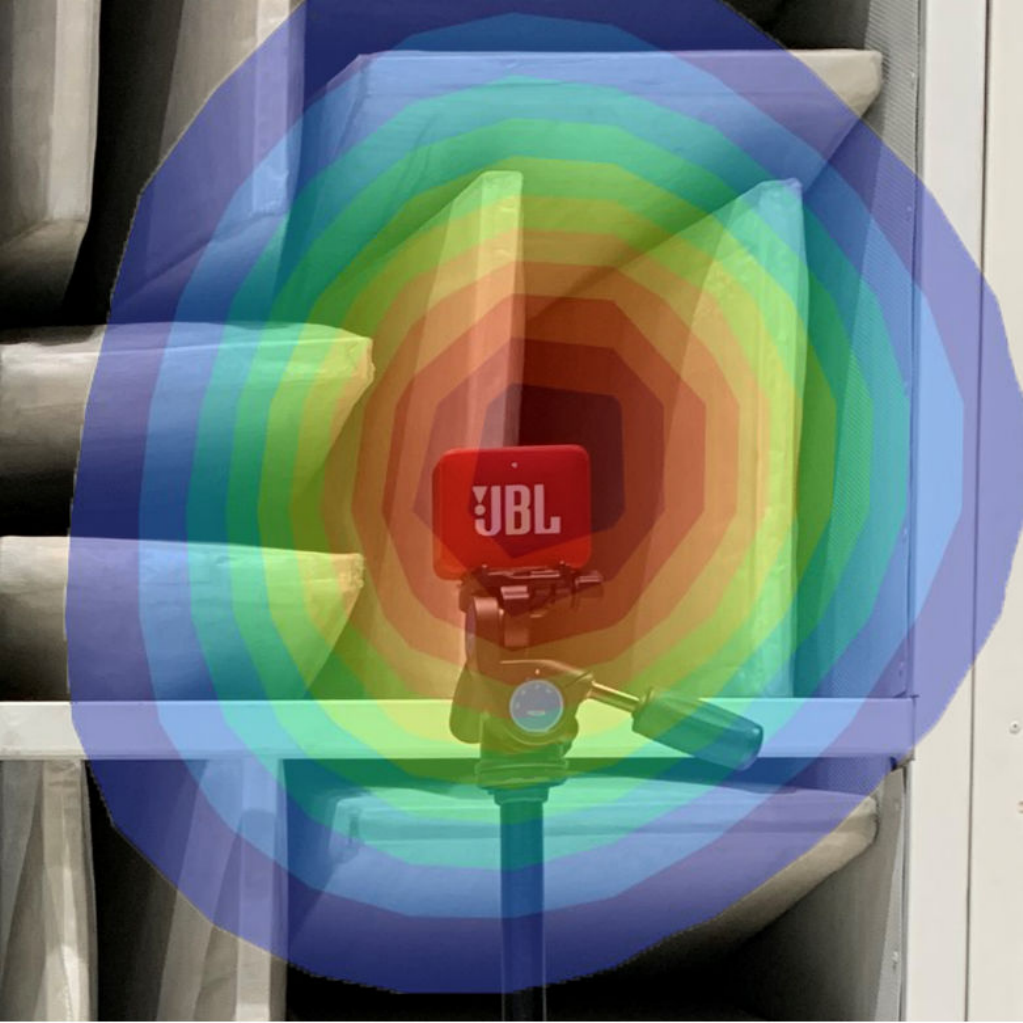}
	}
	\subfigure[]{\includegraphics[width=1.6in]{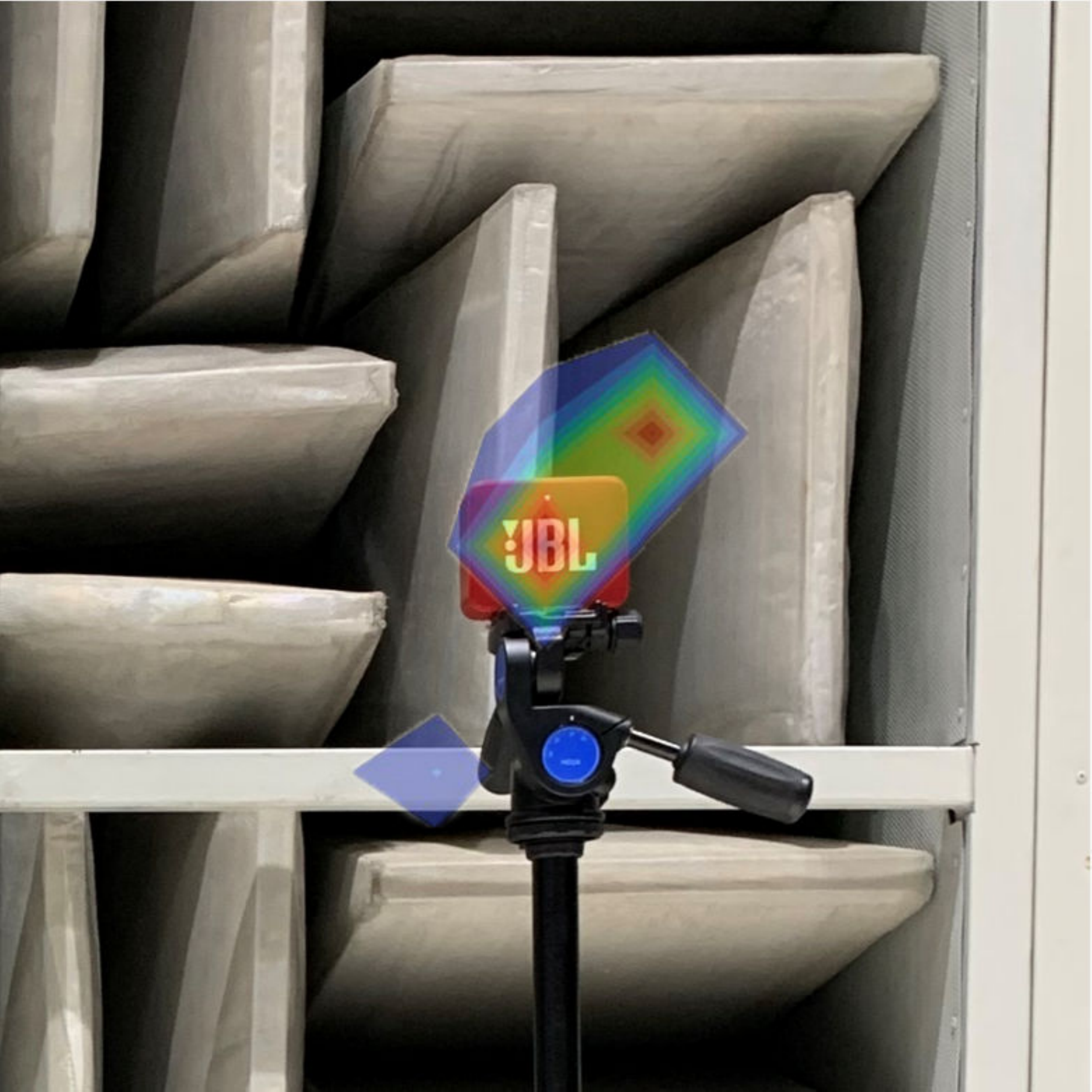}
	}
	\subfigure[]{\includegraphics[width=1.6in]{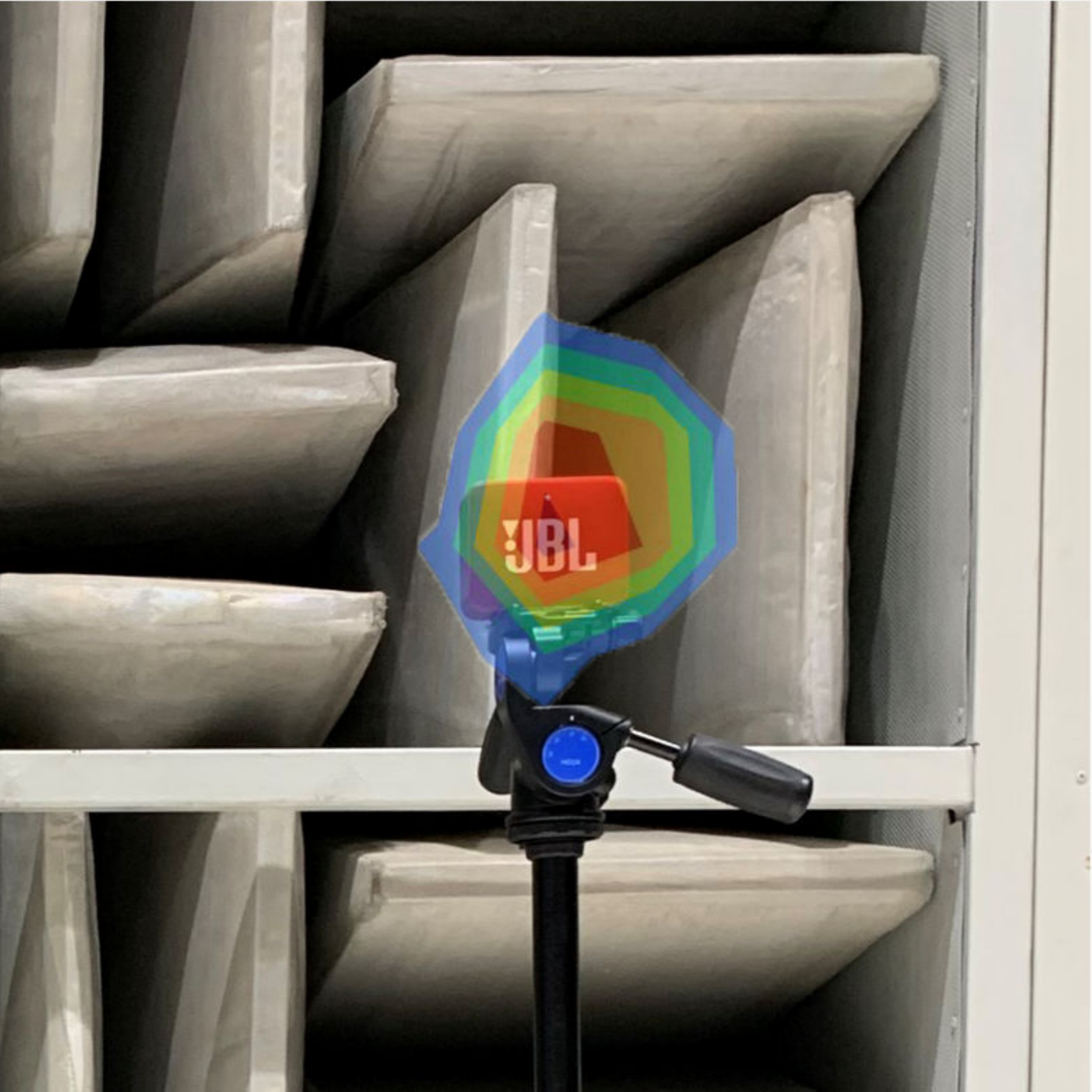}
	}
	\hfil\hfil\\
	\subfigure[]{\includegraphics[width=1.6in]{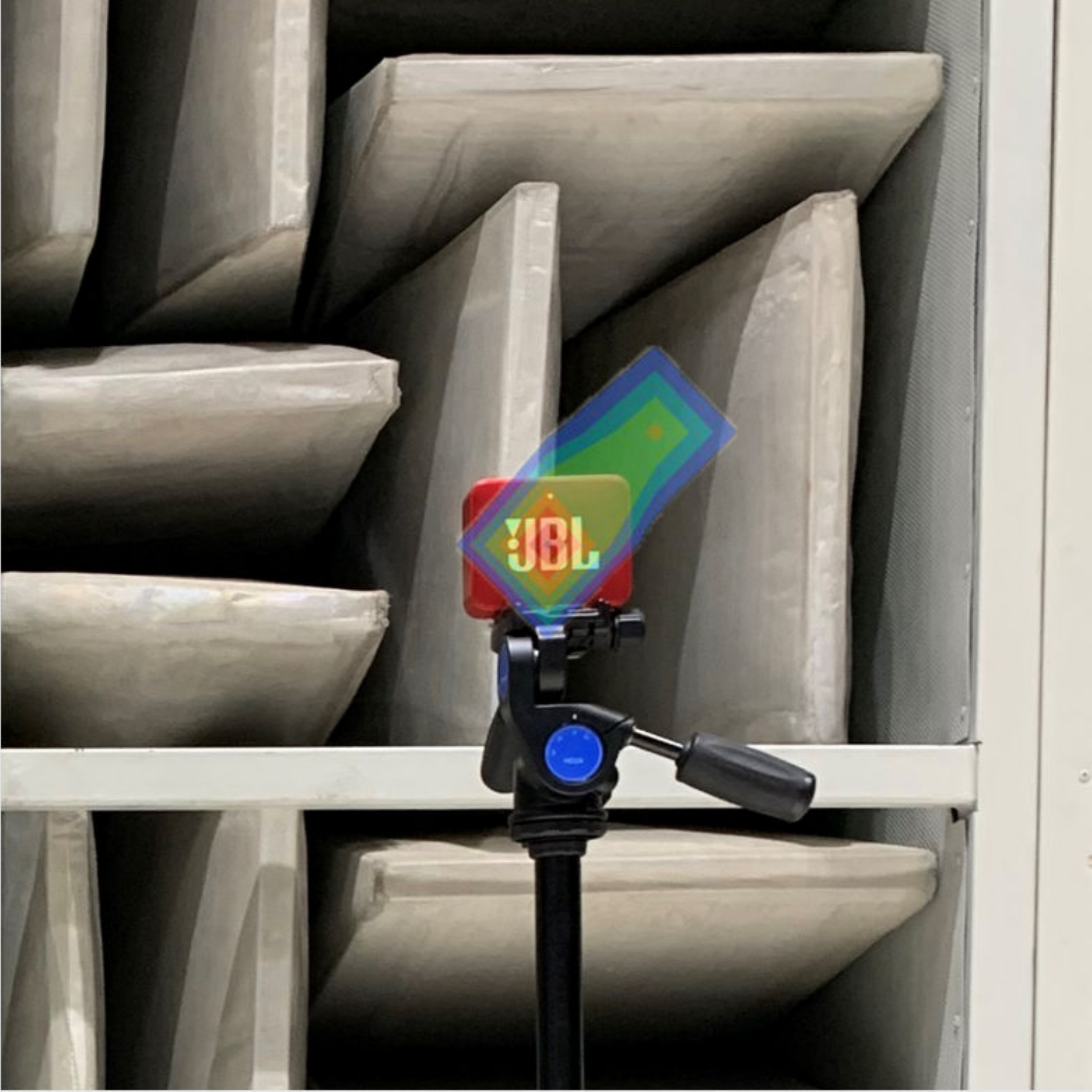}
	}
	\subfigure[]{\includegraphics[width=1.6in]{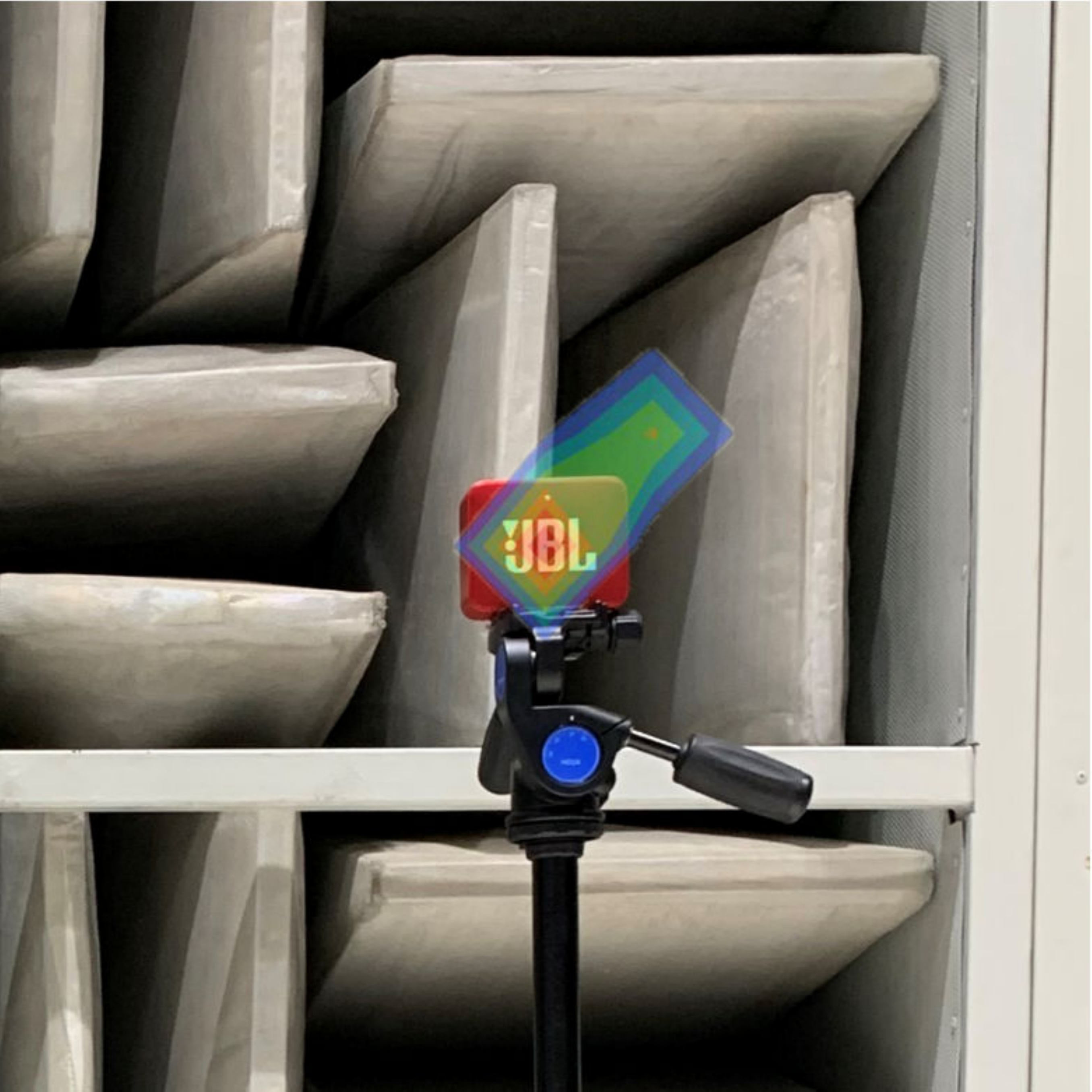}
	}
	\subfigure[]{\includegraphics[width=1.6in]{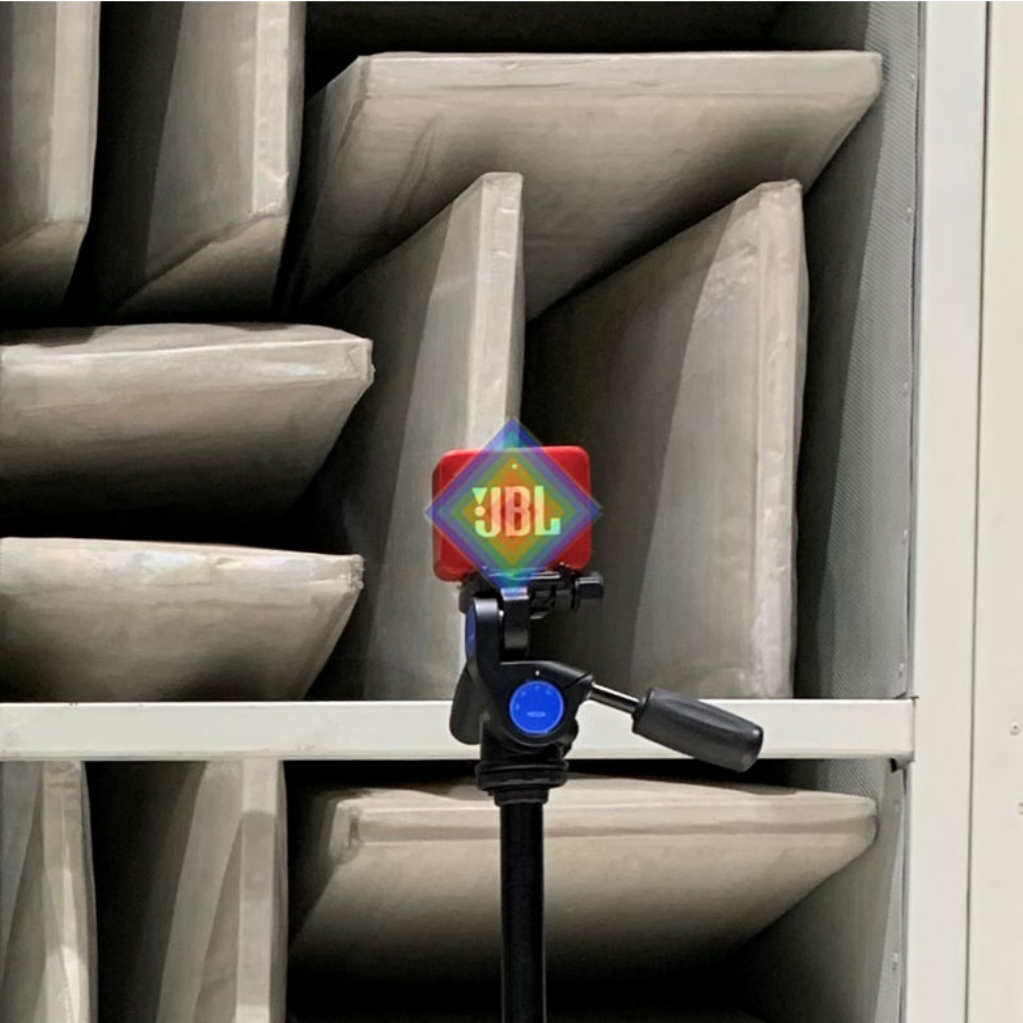}
	}
	\caption{The multi-modal beamforming map of a real-world one-point source by different algorithms: (a) DAS, (b) DAMAS, (c) FFT-NNLS, (d) FFT-DFISTA, (e) DAMAS-FISTA, and (f) DAMAS-FISTA-Net.}
	\label{fig8}
\end{figure*}

Table \ref{table1} and Table \ref{table2} summarize the performance of the different methods for the one-point and two-point source cases, respectively. As may be observed, as compared to the DAMAS, the proposed DAMAS-FISTA always achieves better performance and obtains a higher efficiency. The results clearly illustrate the lower computational complexity of the network methods, although it is clear that these also have a higher bias. It should also be noted that these methods cannot generate the beamforming map directly, and, as noted, may only be used in the one-point source case. In contrast, the proposed DAMAS-FISTA-Net consistently achieves the fastest runtime performance, while having an acceptable bias, clearly indicating the computational advantage of the proposed network. 

Fig.\ref{fig5} and Fig.\ref{fig6} show the corresponding beamforming maps. As may be seen, although the DAS has a low bias and an acceptable runtime, the energy of its beamforming map is not concentrated, resulting in a low-resolution imaging result. As for the FFT-based methods, as mentioned above, the shift-invariant PSF assumption will result in a degree of shift-variance, yielding a larger bias in the estimated location. Although the resolution of DAMAS is high, it may be seen to require a considerable amount of computations to form the imaging. Conversely, the DAMAS-FISTA-Net is able to construct the high-resolution beamforming map efficiently, making it a good candidate for real-time imaging.

\begin{figure*}[!t]
	\centering
	\subfigure[]{\includegraphics[width=2.2in]{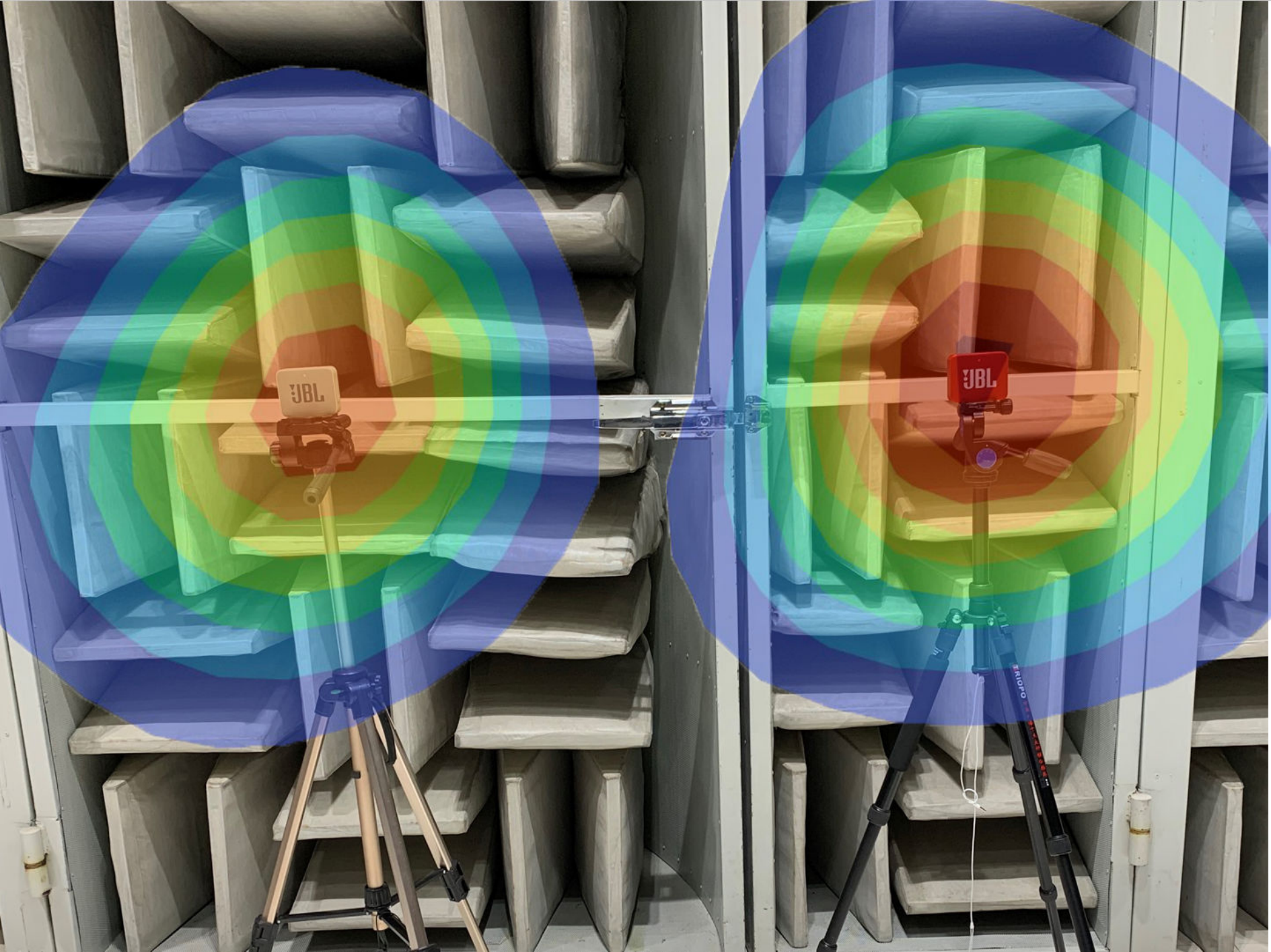}
	}
	\subfigure[]{\includegraphics[width=2.2in]{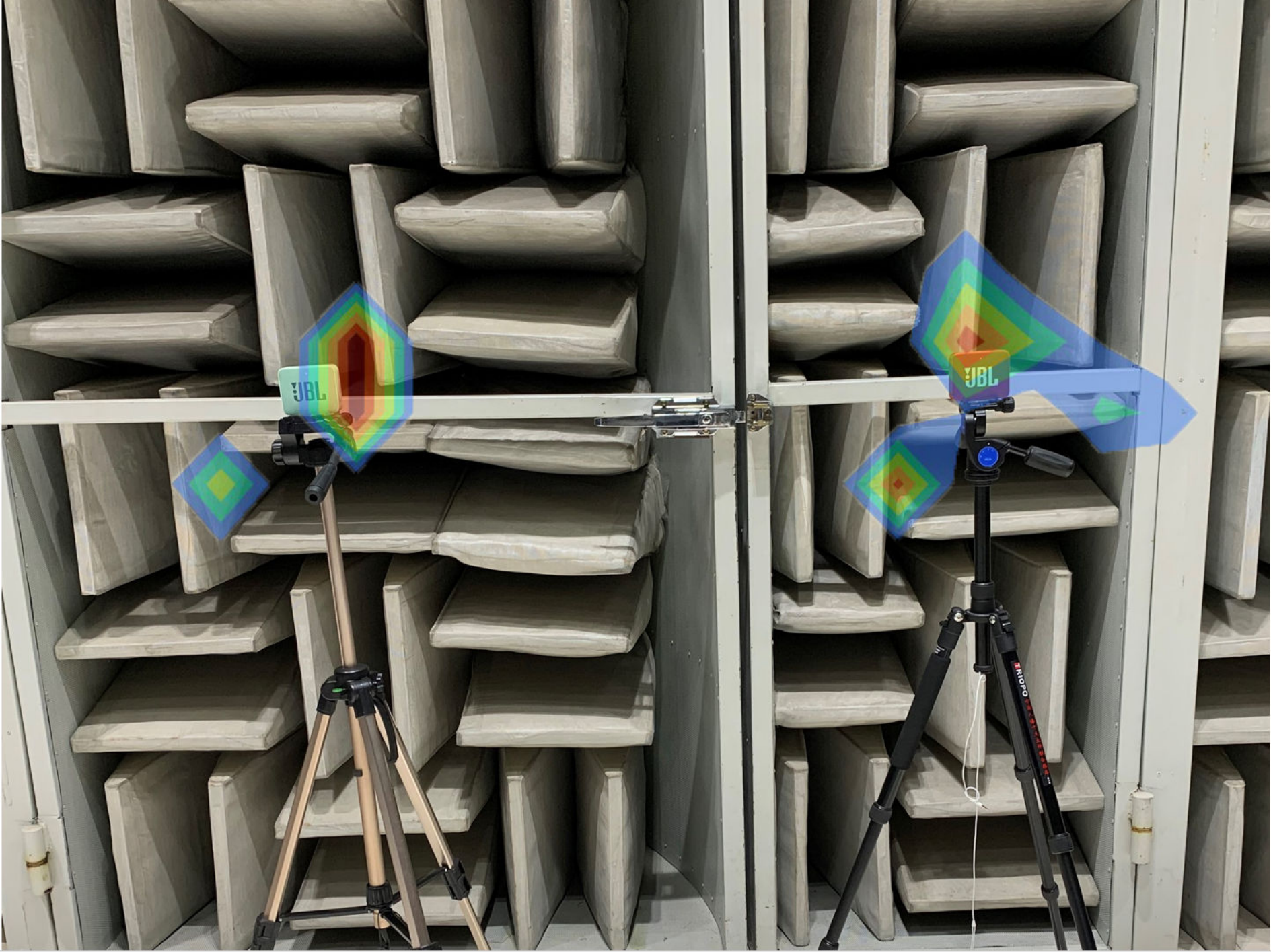}
	}
	\subfigure[]{\includegraphics[width=2.2in]{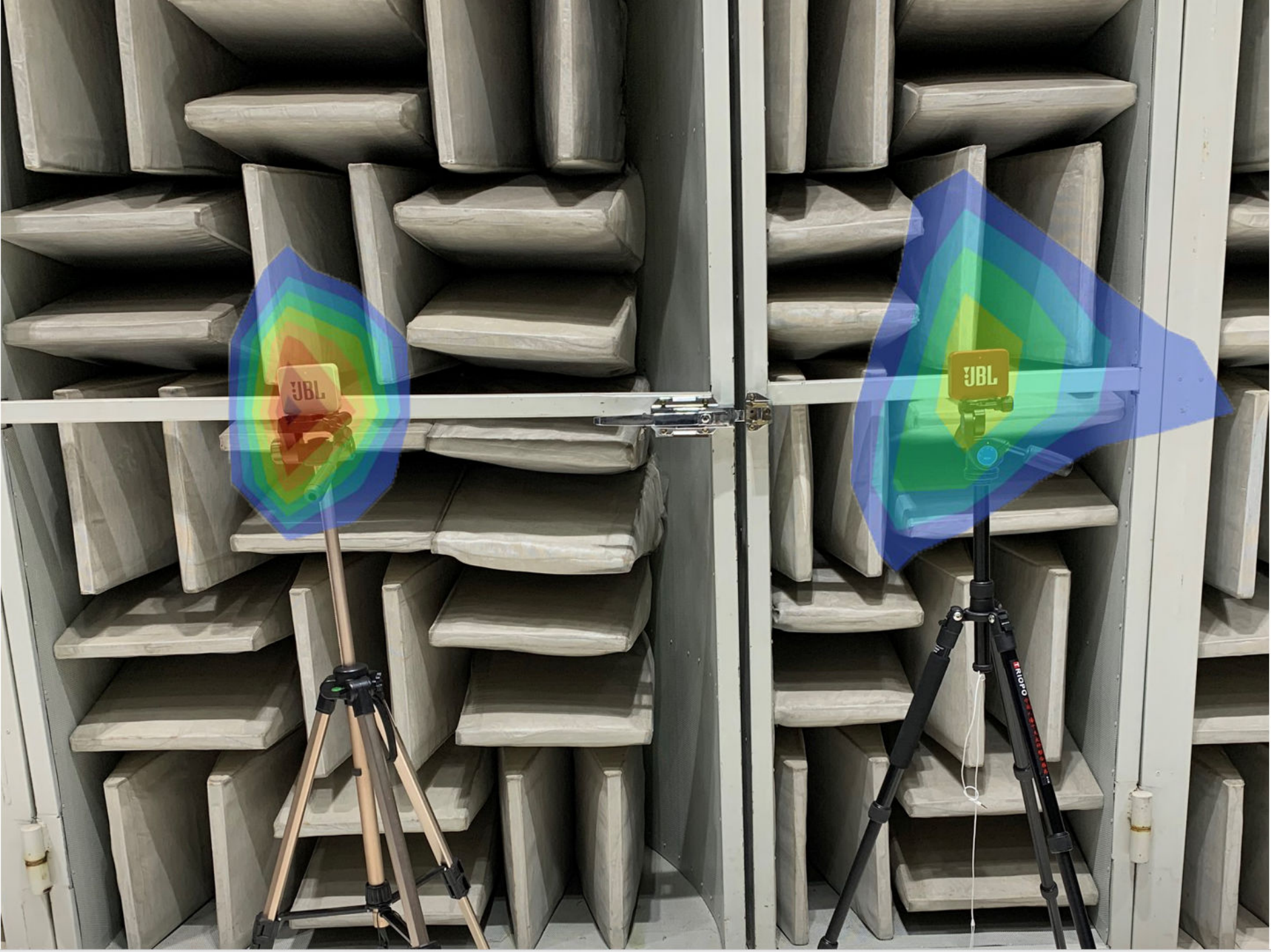}
	}
	\hfil\hfil\\
	\subfigure[]{\includegraphics[width=2.2in]{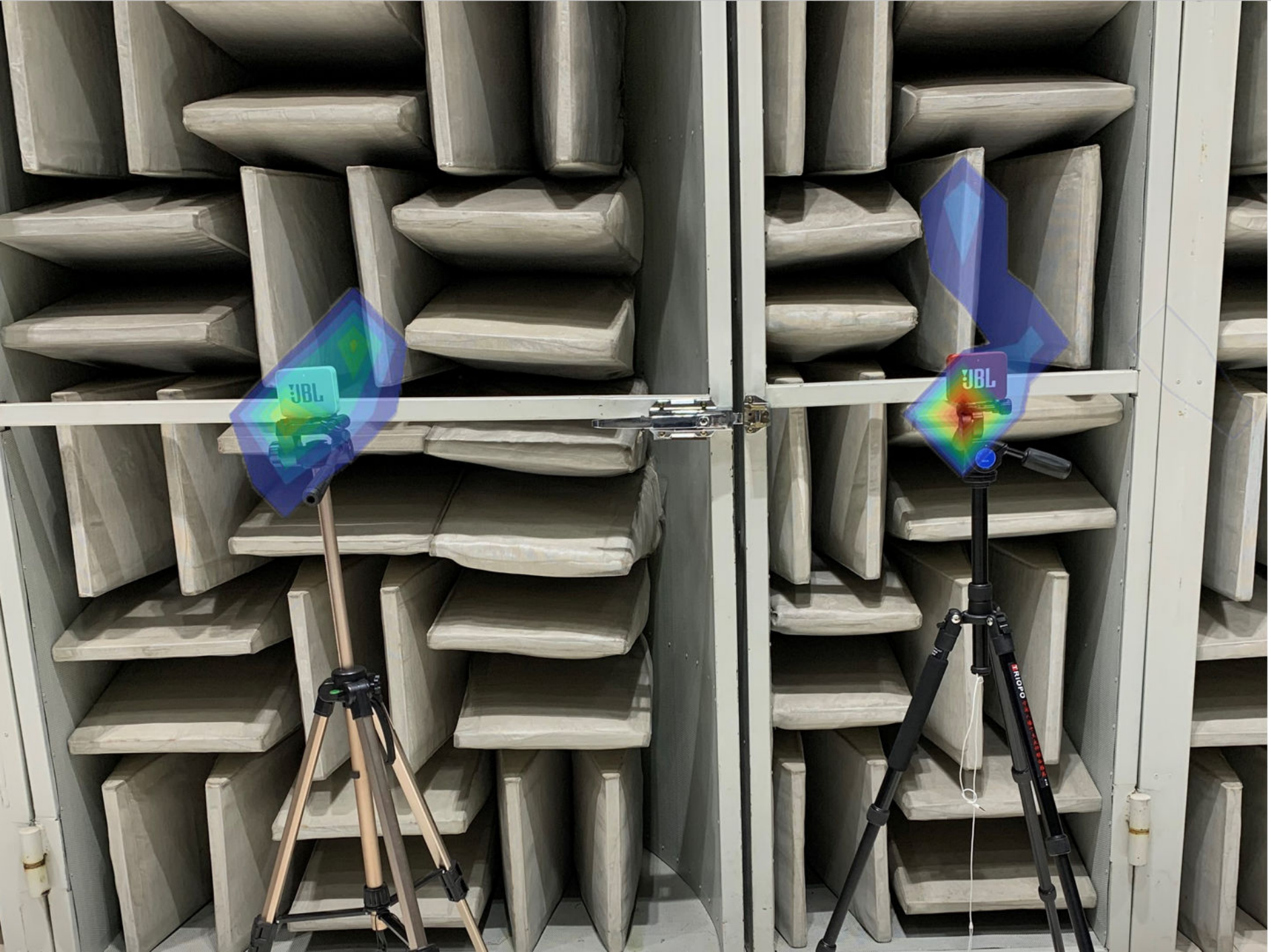}
	}
	\subfigure[]{\includegraphics[width=2.2in]{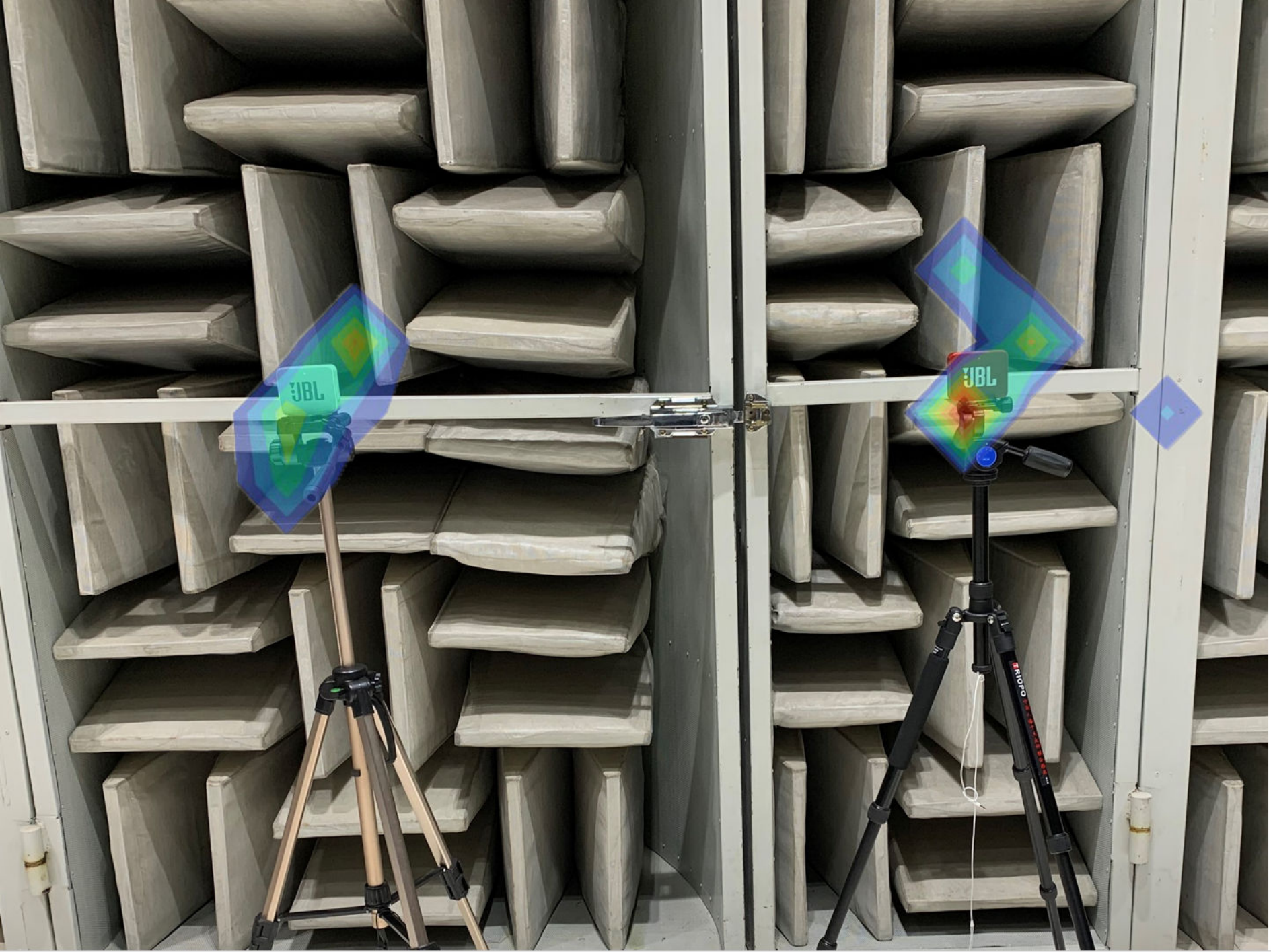}
	}
	\subfigure[]{\includegraphics[width=2.2in]{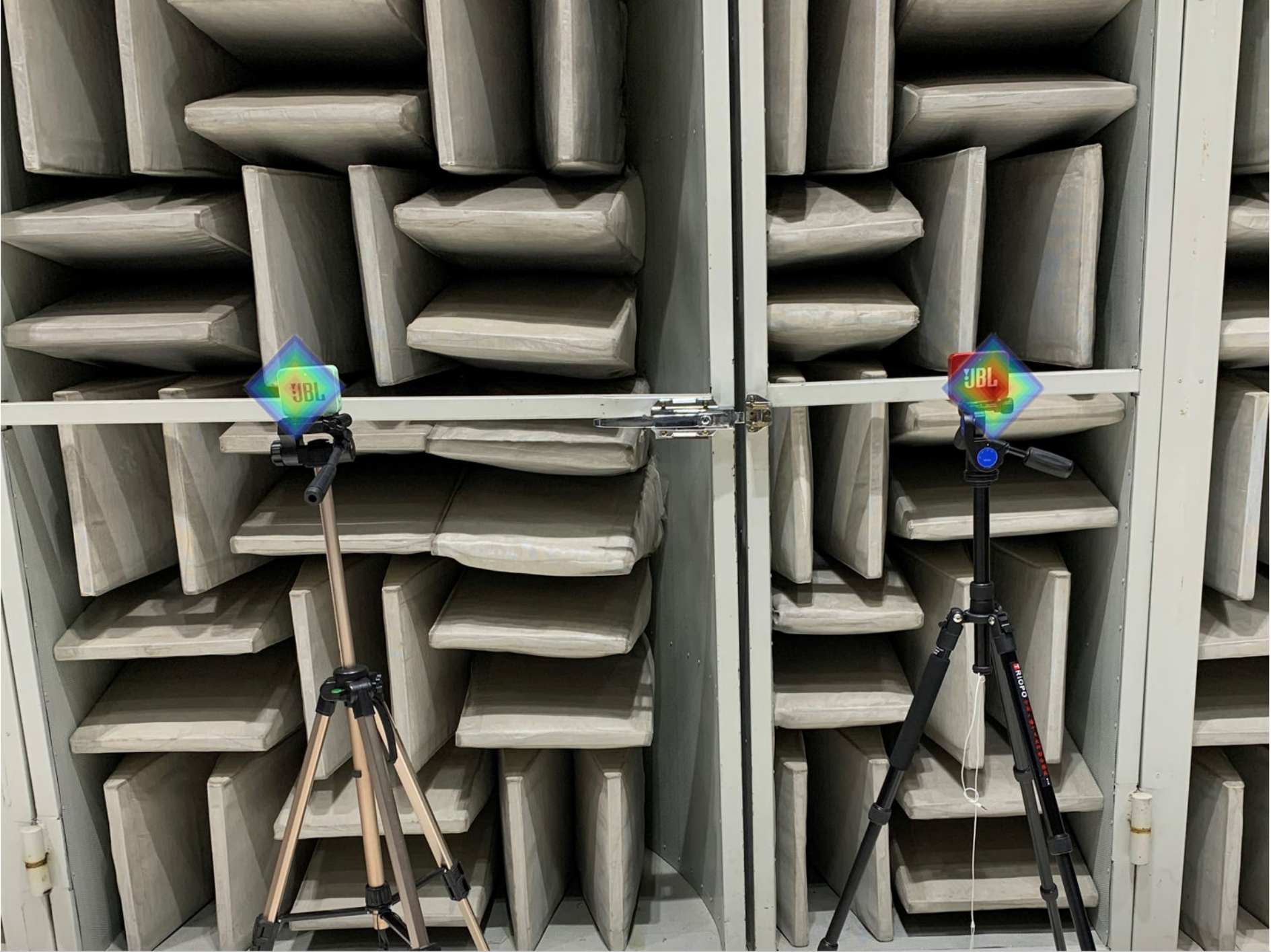}
	}
	\caption{The multi-modal beamforming map of a real-world two-point source by different algorithms: (a) DAS, (b) DAMAS, (c) FFT-NNLS, (d) FFT-DFISTA, (e) DAMAS-FISTA, and (f) DAMAS-FISTA-Net.}
	\label{fig9}
\end{figure*}

\subsection{Real-World Data} \label{dis1}

We proceed to validate our method using real-world experiments. Generally, in real-world scenarios, there is not adequate data with precise labels to allow for reliable network training. Fortunately, our network may learn from the simulated dataset, allowing it to be directly applied to the real-world data. In this section, we evaluate the generalization ability of the DAMAS-FISTA-Net by applying the learned net from the simulated one-point and two-point dataset to real-world one-point and two-point measurement data\footnote{The real-world data will be available online upon publication.}, respectively. This experiment is performed in an anechoic chamber to reduce the effect of noise.

\subsubsection{One-point real source setting}
As shown in Fig.\ref{fig7} (a), we consider a one-point source at a distance of $z = 2.5$ m from the microphone array, having a sampling frequency of 51200 Hz. The signals of the source are generated via a Bluetooth speaker transmitting a 2000 Hz tone, collected by a spiral microphone array with 56 sensors (their coordinates are the same as the array in the simulation).

\subsubsection{Two-point real source setting}
As shown in Fig.\ref{fig7} (b), we consider a two-point source at a distance of $z = 2.5$ m from the microphone array, using the same experiment setup as in the one-point case.

As can be seen in Table \ref{table3} and Table \ref{table4}, the reference deep network-based methods yield a large estimated deviation due to their poor generalization ability, whereas the model-based methods basically generate a low-resolution beamforming map and run slower. Conversly, the proposed DAMAS-FISTA-Net can efficiently form a high-resolution beamforming map. Notably, our network is able to break down the barriers between the simulated and the real-world data, i.e., it only learns from the simulated dataset, and is then able to perform well also on the  real-world data, demonstrating the network's superior generalization ability. Furthermore, the multi-modal beamforming map fused with optical images are shown in Fig.\ref{fig8} and Fig.\ref{fig9}, showing the practical potential of our proposed network.

\begin{table}[!t]
	\centering
	\caption{The experimental indicators by different algorithms for the real-world one-point case}
	\label{table3}
	\begin{tabular}{ccccc}
		\toprule
		Methods          & $R(\alpha)$ $\downarrow$ & $\Delta L$ $\downarrow$& Time $\downarrow$\\ \midrule
		DAS     \cite{van1988beamforming}         &   -0.0133            &        0                    &     0.0748 s         \\ 
		DAMAS     \cite{brooks2006deconvolution}       &      -4.4425         &           0                 &      2.3775 s       \\ 
		FFT-NNLS     \cite{ehrenfried2007comparison} & -3.3147 & 0 & 0.5355 s  \\ 
		FFT-DFISTA    \cite{ding2022high}    &  -4.5839             &             0               &    2.8353 s          \\ \midrule
		Acoustic-AlexNet \cite{reiter2017machine} & -             &       0.4243                     &      0.0784 s        \\ 
		Acoustic-ResNet \cite{kujawski2019deep} & -             &        1.6733                    &    0.0797 s          \\ 
		Acoustic-Net   \cite{zhou2022acoustic}  & -             &       0.5439                     &     0.0565 s         \\ \midrule
		DAMAS-FISTA (Ours)    &   -4.9180            &               0             &   1.5821 s           \\ 
		DAMAS-FISTA-Net (Ours) &    \textbf{-5.7627}          &         0                   &    \textbf{0.0072 s}          \\ \bottomrule
	\end{tabular}
\end{table}

\begin{table}[!t]
	\centering
	\caption{The experimental indicators by different algorithms for the real-world two-point case}
	\label{table4}
	\begin{tabular}{ccccc}
		\toprule
		Methods          & $R(\alpha)$ $\downarrow$ & $\Delta L$ $\downarrow$& Time $\downarrow$\\ \midrule
		DAS     \cite{van1988beamforming}         &        1.0571       &         0.1000                   &     0.0734 s         \\ 
		DAMAS     \cite{brooks2006deconvolution}       &    -3.0036          &          0.1207                 &    13.7383 s       \\ 
		FFT-NNLS     \cite{ehrenfried2007comparison} & -1.5530  &   0 & 0.6463 s \\ 
		FFT-DFISTA    \cite{ding2022high}    &     -3.1235
		         &     0                       &    4.0994 s          \\ \midrule
		DAMAS-FISTA (Ours)    &       -3.4238        &    0.0707                      &      2.0231 s       \\ 
		DAMAS-FISTA-Net (Ours) &      \textbf{-4.6702}         &    0.1000                     &     \textbf{0.0070 s}        \\ \bottomrule
	\end{tabular}
\end{table}


\begin{table*}[!t]
	\centering
	\caption{Generalization Studies}
	\label{table5}
	\resizebox{1.0\textwidth}{!}{
	\begin{tabular}{ccccccc}
		\toprule
		Methods          & Training & Validation & $R(\alpha)$ $\downarrow$ & $\Delta L$ $\downarrow$&Time $\downarrow$\\ \midrule
		DAMAS-FISTA-Net$_{s1\rightarrow s1}$  &     Simulated one-point data        &    Simulated one-point data                     &   \textbf{-6.5235} & \textbf{0.0028} & 0.0125 s         \\ 
		DAMAS-FISTA-Net$_{s2\rightarrow s1}$  &     Simulated two-point data                                &      Simulated one-point data & -6.0999 & 0.0049 &   \textbf{0.0123 s}\\ \midrule
		DAMAS-FISTA-Net$_{s1\rightarrow s2}$  &      Simulated one-point data       &         Simulated two-point data                 &    -5.2107   & \textbf{0.0371}   & 0.0126 s \\ 
		DAMAS-FISTA-Net$_{s2\rightarrow s2}$  &      Simulated two-point data         &     Simulated two-point data                     &  \textbf{-5.5567}  & 0.0584 & \textbf{0.0124 s}    \\ \midrule
		DAMAS-FISTA-Net$_{s1\rightarrow r1}$  &      Simulated one-point data       &       Real-world one-point data                  &     -5.7627 & 0  & \textbf{0.0072 s}      \\ 
		DAMAS-FISTA-Net$_{s2\rightarrow r1}$  &      Simulated two-point data         &      Real-world one-point data                   &   \textbf{-5.7849} & 0 &   0.0074 s   \\ \midrule
		DAMAS-FISTA-Net$_{s1\rightarrow r2}$  &      Simulated one-point data       &     Real-world two-point data                    &    -4.0919 & 0.100 & 0.0078 s       \\ 
		DAMAS-FISTA-Net$_{s2\rightarrow r2}$  &      Simulated two-point data        &       Real-world two-point data                  &     \textbf{-4.6702} & 0.100  & \textbf{0.0070 s}   \\  \bottomrule
	\end{tabular}
	}
\end{table*}

\begin{table*}[!t]
	\centering
	\caption{The experimental indicators by different algorithms for the simulated one-point case under different noise levels}
	\label{table6}
	\resizebox{1.0\textwidth}{!}{
	\begin{tabular}{ccccccccccc}
		\toprule
		Noise levels            & Indicators & DAS \cite{van1988beamforming} & DAMAS \cite{brooks2006deconvolution}& FFT-NNLS \cite{ehrenfried2007comparison}& FFT-DFISTA \cite{ding2022high} & Acoustic-AlexNet \cite{reiter2017machine} & Acoustic-ResNet \cite{kujawski2019deep} & Acoustic-Net   \cite{zhou2022acoustic} & DAMAS-FISTA (Ours) & \multicolumn{1}{l}{DAMAS-FISTA-Net (Ours)} \\ \midrule
		\multirow{3}{*}{-10 dB} & $R(\alpha)$ $\downarrow$         &   -0.2847  &   -6.4321    &    -2.2467     &     -4.0855      &      -              & - & - &  \textbf{-6.5271} &-6.5161                       \\
		& $\Delta L$ $\downarrow$          &   0  &   0    &   0.0983       &    0.1157      &     0              & 0.0290 & 1.4609 & 0 &0.0028                     \\  
		& Time  $\downarrow$     &  0.0658 s   &    10.7178 s   &   0.3559 s      &    5.4640 s       &       0.0722 s             &  0.0705 s & 0.0579 s &  1.2101 s& \textbf{0.0123 s}                       \\ \midrule
		\multirow{3}{*}{0 dB}   & $R(\alpha)$ $\downarrow$         &   -0.2966  &   -6.5740    &     -2.2705     &         -4.1005                        &            -                            &           -           & -& \textbf{-6.6252} & -6.5228                     \\
		& $\Delta L$ $\downarrow$          &   0  &   0    &     0.0993     &    0.1162                            &        0                                &   0.0288  & 1.3063 & 0 &  0.0028                                   \\
		& Time $\downarrow$      &   0.0661 s  &   11.1752 s    &   0.3646 s       &     5.2969 s                           &         0.0712 s                               &      0.0709 s     & 0.0584
		 s& 1.2234 s &   \textbf{0.0124 s}                              \\ \midrule
		\multirow{3}{*}{10 dB}  & $R(\alpha)$ $\downarrow$         &   -0.2983  &   -6.6102    &  -2.2673       &      -4.1046                          &                -                        &    -   &  -   & \textbf{-6.6410} &   -6.5234                                 \\
		& $\Delta L$ $\downarrow$          &   0  &   0    &  0.1000        &     0.1159                           &         0                               &   0.0288   & 0.7741 & 0 &  0.0028                                   \\
		& Time   $\downarrow$    &  0.0665 s   &    11.5698 s   &    0.3653 s      &      5.1992 s                          &               0.0716 s                         &     0.0715 s     & 0.0588 s& 1.1177 s&   \textbf{0.0124 s}                              \\  \midrule
		\multirow{3}{*}{Noise-free}  & $R(\alpha)$ $\downarrow$         &  -0.2984   &    -6.6220   &      -2.2609     &       -4.1067                          &        -                                &     -         & - & \textbf{-6.6431} &-6.5235                              \\
		& $\Delta L$ $\downarrow$          &  0   &    0   &    0.0987      &    0.1151                            &         0                               &     0.0288        & 0.0231 & 0 &  0.0028                              \\
		& Time   $\downarrow$    &  0.0655 s   &    11.9595 s   &     0.3547 s     &      5.2308 s                          &           0.0721 s                             &   0.0711 s &  0.0588 s & 1.0689 s &     \textbf{0.0125 s}                                  \\ \bottomrule
	\end{tabular}
	}
\end{table*}

\begin{table*}[!t]
	\centering
	\caption{Comparison of different network depths in the DAMAS-FISTA-Net}
	\label{table7}
	\begin{tabular}{cccccc}
		\toprule
		Network Depths & $L=3$ & $L=4$ & $L=5$ & $L=6$ & $L=7$ \\ \midrule
		$R(\alpha)$ $\downarrow$              &  -6.4632    &   -6.5075   &  \textbf{-6.5235}   & -6.5068 &  -6.4744    \\
		$\Delta L$ $\downarrow$              &   0.0097   &  \textbf{0.0027}   &   0.0028  & 0.0057 &  0.0057  \\
		Time    $\downarrow$       &  \textbf{0.0103 s}    &    0.0114 s   &  0.0125 s   &   0.0133 s &  0.0145 s \\ \bottomrule
	\end{tabular}
\end{table*}

\subsection{Discussions}

In the above experiments, the proposed DAMAS-FISTA-Net is able to produce highly accurate results for both simulated and real-world data. Proceeding, we examine the generalization ability of DAMAS-FISTA-Net and its sensitivity to noise. The effects of different network depths are also evaluated.

\subsubsection{Generalization Studies}
As evaluated in Section \ref{dis1}, the proposed network is able to generalize well from simulated data to real-world data. Here, we further test its generalization ability in the multi-source situation, applying the learned network from the simulated one-point source dataset to simulated two-point source data. Table \ref{table5} summarizes the indicators of the different algorithms in constructing the beamforming map of the different situations. Notably, our network can be seen to achieve competitive performance even when the training and validation data are biased (e.g., the network learns from the simulated one-point source dataset but is validated on the two-point source scene), again revealing its robust generalization ability. The reason for this robustness may be due to our hybrid approach, which combines both the model-based scheme and the deep learning framework, allowing the network to inherit relevant domain knowledge. In contrast, the deep network-based methods can only form an estimate for the one-point source situation due to their network output settings, which may severely limit their applicability.

\subsubsection{Sensitivity to Noise}
We next extend the proposed method to process array data with Gaussian white noise in order to demonstrate the robustness of the DAMAS-FISTA-Net. Here, we add -10 dB, 0 dB, and 10 dB Gaussian white noise to the signals in the validation set, respectively. Table \ref{table6} shows the comparison results under different noise levels. As can be observed, the acoustic imaging performance of the network is only mildly influenced by the added noise, e.g., the R\'{e}nyi entropy is only increased by 0.0074 by the addition of noise with noise levels -10 dB. Notably, we do not need to retrain the DAMAS-FISTA-Net for different noise levels, implying that the learned network has a certain robustness to the influence of noise.

\subsubsection{Evaluation for Different Network Depths}

Finally, we investigate the effect of different network depths on the network's performance. The testing beamforming results in Table \ref{table7} clearly show that within a certain number of layers, using more deeper layers could effectively improve the imaging quality, although at the cost of a larger computational overhead. However, it can also be seen that adding more layers to the network, e.g., when the network depths, $L$, is larger than 5, may also degrade the network's performance, which may be due to the overfitting of the dataset.

\section{Conclusion} \label{s6}

In this paper, we propose a computationally efficient model-based acoustic beamforming method, termed the DAMAS-FISTA. Inspired by the model-based deep learning theory, we further propose a hybrid structured model-based deep learning network for real-time imaging, dubbed the DAMAS-FISTA-Net, which is able to process the raw measurement data directly. In order to achieve real-time execution and generalizable performance, the network exploits the structure of the model-based estimator, enabling the network to inherit physical knowledge and learn the underlying physical properties of the acoustic scene. The performance of the proposed network is via extensive simulated and real-world experiments shown to offer preferable performance as compared to alternative approaches, while also indicating the potential practical applicability of the estimator.

\balance
\bibliographystyle{IEEEtran}
\bibliography{ref}
\end{document}